\documentclass{aa}

\usepackage{graphicx}
\usepackage{txfonts}
\usepackage{graphics,graphicx}
\usepackage{amsmath}	
\usepackage{amssymb}	
\usepackage{savesym}
\usepackage{enumitem}
\usepackage{mathrsfs,dsfont}
\usepackage{multirow,multicol}
\usepackage{captcont}
\usepackage{float}
\usepackage[export]{adjustbox}
\usepackage{booktabs}
\usepackage{epstopdf}
\usepackage{soul}
\usepackage{xcolor}
\usepackage[colorlinks=true,linkcolor=blue,citecolor=blue,urlcolor=black]{hyperref}
\usepackage{orcidlink} 
\usepackage{cleveref}
\usepackage{verbatim}
\usepackage{makecell}

\def\be{\begin{equation}}
\def\ee{\end{equation}}
\def\kms{{\rm \,km\,s^{-1}}}
\def\s{{\rm \,s}}

\def\Gyr{{\rm \,Gyr}}

\def\Mpc{{\rm \,Mpc}}
\def\kpc{{\rm \,kpc}}
\def\cm{{\rm \,cm}}
\def\erg{{\rm \,erg}}
\def\eV{{\rm \,eV}}
\def\keV{{\rm \,keV}}

\def\msun{{\,M_\odot}}

\newcommand{\chandra}{{\em Chandra}}
\newcommand{\xmm}{{\em XMM-Newton}}
\newcommand{\suzaku}{{\em Suzaku}}

\newcommand{\orcid}[1]{\orcidlink{#1}}

\graphicspath{{figures/}}

\begin{document}

\title{Mapping the Perseus Galaxy Cluster with XRISM}

\subtitle{Gas Kinematic Features and their Implications for Turbulence}

\author{Congyao Zhang\orcid{0000-0001-5888-7052}\thanks{\email{cyzhang@astro.uchicago.edu}}\inst{\ref{aff1},\ref{aff2}}
\and Irina Zhuravleva\orcid{0000-0001-7630-8085}\inst{\ref{aff2}}
\and Annie Heinrich\orcid{0000-0002-7726-4202}\inst{\ref{aff2}}
\and Elena Bellomi\orcid{0000-0001-6411-3686}\inst{\ref{aff3}}
\and Nhut Truong\orcid{0000-0003-4983-0462}\inst{\ref{aff4},\ref{aff5},\ref{aff6}}
\and John ZuHone\orcid{0000-0003-3175-2347}\inst{\ref{aff3}}
\and Eugene Churazov\orcid{0000-0002-0322-884X}\inst{\ref{aff7},\ref{aff8}}
\and Megan E.\ Eckart\orcid{0000-0003-3894-5889}\inst{\ref{aff9}}
\and Yutaka Fujita\orcid{0000-0003-0058-9719}\inst{\ref{aff10}}
\and Julie Hlavacek-Larrondo\orcid{0000-0001-7271-7340}\inst{\ref{aff101}}
\and Yuto Ichinohe\orcid{0000-0002-6102-1441}\inst{\ref{aff102}}
\and Maxim Markevitch\orcid{0000-0003-0144-4052}\inst{\ref{aff5}}
\and Kyoko Matsushita\orcid{0000-0003-2907-0902}\inst{\ref{aff103}}
\and Fran\c{c}ois Mernier\orcid{0000-0002-7031-4772}\inst{\ref{aff104}}
\and Eric D.\ Miller\orcid{0000-0002-3031-2326}\inst{\ref{aff13}}
\and Koji Mori\orcid{0000-0002-0018-0369}\inst{\ref{aff14}}
\and Hiroshi Nakajima\orcid{0000-0001-6988-3938}\inst{\ref{aff20}}
\and Anna Ogorzalek\orcid{0000-0003-4504-2557}\inst{\ref{aff21},\ref{aff5},\ref{aff6}}
\and Frederick S.\ Porter\orcid{0000-0002-6374-1119}\inst{\ref{aff5}}
\and Ay\c seg\"ul T\"umer\orcid{0000-0002-3132-8776}\inst{\ref{aff4},\ref{aff5},\ref{aff6}}
\and Shutaro Ueda\orcid{0000-0001-6252-7922}\inst{\ref{aff22},\ref{aff23},\ref{aff24}}
\and Norbert Werner\orcid{0000-0003-0392-0120}\inst{\ref{aff1}}
}

\institute{Department of Theoretical Physics and Astrophysics, Masaryk University, Brno 61137, Czechia\label{aff1}
\and Department of Astronomy and Astrophysics, The University of Chicago, Chicago, IL 60637, USA\label{aff2}
\and Center for Astrophysics | Harvard-Smithsonian, MA 02138, USA\label{aff3}
\and Center for Space Sciences and Technology, University of Maryland, Baltimore County (UMBC), Baltimore, MD, 21250 USA\label{aff4}
\and NASA / Goddard Space Flight Center, Greenbelt, MD 20771, USA\label{aff5}
\and Center for Research and Exploration in Space Science and Technology, NASA / GSFC (CRESST II), Greenbelt, MD 20771, USA\label{aff6}
\and Max Planck Institute for Astrophysics, Karl-Schwarzschild-Str. 1, D-85741 Garching, Germany\label{aff7}
\and Space Research Institute (IKI), Profsoyuznaya 84/32, Moscow 117997, Russia\label{aff8}
\and Lawrence Livermore National Laboratory, Livermore, CA 94550, USA\label{aff9}
\and Department of Physics, Tokyo Metropolitan University, Tokyo 192-0397, Japan\label{aff10}
\and D\'epartement de Physique, Universit\'e de Montr\'eal, Succ. Centre-Ville, Montr\'eal, Qu\'ebec H3C 3J7, Canada\label{aff101}
\and RIKEN Nishina Center, Saitama 351-0198, Japan\label{aff102}
\and Faculty of Physics, Tokyo University of Science, Tokyo 162-8601, Japan\label{aff103}
\and IRAP, CNRS, Université de Toulouse, CNES, UT3-UPS, Toulouse, France\label{aff104}
\and Kavli Institute for Astrophysics and Space Research, Massachusetts Institute of Technology, Cambridge, MA 02139, USA\label{aff13}
\and Faculty of Engineering, University of Miyazaki, Miyazaki 889-2192, Japan\label{aff14}
\and College of Science and Engineering, Kanto Gakuin University, Kanagawa 236-8501, Japan\label{aff20}
\and Department of Astronomy, University of Maryland, College Park, MD 20742, USA\label{aff21}
\and Faculty of Mathematics and Physics, Kanazawa University, Kanazawa, Ishikawa, 920-1192, Japan\label{aff22}
\and Advanced Research Center for Space Science and Technology, Kanazawa University, Kanazawa, Ishikawa, 920-1192, Japan\label{aff23}
\and Academia Sinica Institute of Astronomy and Astrophysics (ASIAA), Taipei, 106319, Taiwan\label{aff24}
}

\date{Received xxx; accepted xxx}

  \abstract{In this paper, we present extended gas kinematic maps of the Perseus cluster by combining five new XRISM/Resolve pointings observed in 2025 with four Performance Verification datasets from 2024, totaling 745 ks net exposure. To date, Perseus remains the only cluster that has been extensively mapped out to $\simeq0.7r_{2500}$ by XRISM/Resolve, while simultaneously offering sufficient spatial resolution to resolve gaseous substructures driven by mergers and active galactic nucleus (AGN) feedback. Our observations cover multiple radial directions and a broad range of dynamical scales, enabling us to characterize the kinematic properties of the intracluster medium up to the scale of $\sim500\kpc$. In the measurements, we detect high velocity dispersions ($\simeq300\kms$) in the eastern region of the cluster, spatially coincident with the extended X-ray surface brightness excess and corresponding to a nonthermal pressure fraction of $\simeq7-13\%$. The velocity field outside the AGN-dominant region can be effectively described by a single, large-scale kinematic driver based on the velocity structure function, which statistically favors an energy injection scale of at least a few hundred kpc. The estimated turbulent dissipation energy is comparable to the gravitational potential energy released by a recent merger, implying a significant role of turbulent cascade in the merger energy conversion. In the bulk velocity field, we observe a dipole-like pattern along the east-west direction with an amplitude of $\simeq\pm200-300\kms$, indicating rotational motions induced by the recent merger event. This feature constrains the viewing direction to $\simeq30^\circ-50^\circ$ relative to the normal of the merger plane. Our hydrodynamic simulations suggest that Perseus has experienced at least two energetic mergers since redshift $z\sim1$, the most recent of which is associated with the radio galaxy IC310, in agreement with recent SRG/eROSITA findings. This study showcases exciting scientific opportunities for future missions with high-resolution spectroscopic capabilities (e.g., HUBS, LEM, and NewAthena). }

   \keywords{ Galaxies: clusters: individual: Perseus -- Galaxies: clusters: intracluster medium -- Methods: observational -- Techniques: imaging spectroscopy
              -- Turbulence -- X-rays: galaxies: clusters
               }

   \maketitle
%

\section{Introduction}

The Perseus cluster is a prototypical cool-core galaxy cluster system that hosts a variety of remarkable gaseous structures (e.g., shocks, cold fronts, bubbles, and turbulent fluctuations), which have been extensively studied with all major X-ray telescopes \citep[e.g.,][]{Forman1972,Schwarz1992,Fabian2000,Churazov2003,Simionescu2012,Zhuravleva2015,Hitomi2016_Nature} as well as other wavelength observations \citep[e.g.,][]{Salome2006,Canning2014,Gendron-Marsolais2020,vanWeeren2024,Cuillandre2025}. It has been widely regarded as a textbook example of radio-mode active galactic nucleus (AGN) feedback and merger-driven sloshing  -- two major physical processes that shape cluster gas atmospheres (a.k.a., intracluster medium, or ICM), especially the central region (see \citealt{Markevitch2007,Fabian2012,ZuhoneRoediger2016} for reviews). In theoretical studies, Perseus often serves as a diagnostic benchmark for models of AGN feedback and microphysics of the ICM implemented in numerical simulations \citep[e.g.,][]{Reynolds2005,Li2014,Zhang2022,Ewart2024}.

The dynamical state and assembly history of the Perseus cluster have long been investigated \citep[e.g.,][]{Simionescu2012,Kang2024,Churazov2025}, as they are essential for understanding the formation of X-ray and radio structures (e.g., cold front edges at different radii, co-existence of radio halos of various types), the dynamical impact of the AGN feedback throughout the entire cool core, and the secular evolution of member galaxies. Direct gas velocity measurements are expected to offer critical insights into these processes (see \citealt{Simionescu2019} for a review).

Hitomi pioneered the high-resolution, non-dispersive imaging spectroscopy of the hot cluster atmosphere in Perseus, detecting gas motions with the velocity dispersion $\simeq80-220\kms$ within the radius of $\simeq60\kpc$, $\sim1/3$ of the cool-core radius \citep{Hitomi2016_Nature,Hitomi2018_Velocity}. The recently-launched XRISM observatory is equipped with the Resolve X-ray microcalorimeter, which offers a high energy resolution of $\simeq4.5\eV$ (full width at half maximum, FWHM; \citealt{Tashiro2021,Ishisaki2022}). It is designed to carry forward the mission initiated by the short-lived Hitomi, reopening the window into the high-precision ICM kinematics.

The X-ray microcalorimeter enables spatial mapping of the ICM kinematics, which is critical for characterizing mass and energy flows in galaxy clusters and essential for understanding cluster assembly processes and the plasma physics of the ICM \citep[e.g.,][]{Kitayama2014,Zhang2024}. However, Resolve's moderate angular resolution of $\simeq1.3'$ (half-power diameter, HPD) and relatively small effective area (i.e., $\simeq200\cm^2$ at $6\keV$) practically require the mapping target to be nearby and sufficiently bright. Perseus, being the brightest cluster in the X-ray sky and located at a low redshift, is thus an optimal target for the goal.

\begin{figure}
\centering
\includegraphics[width=0.95\linewidth]{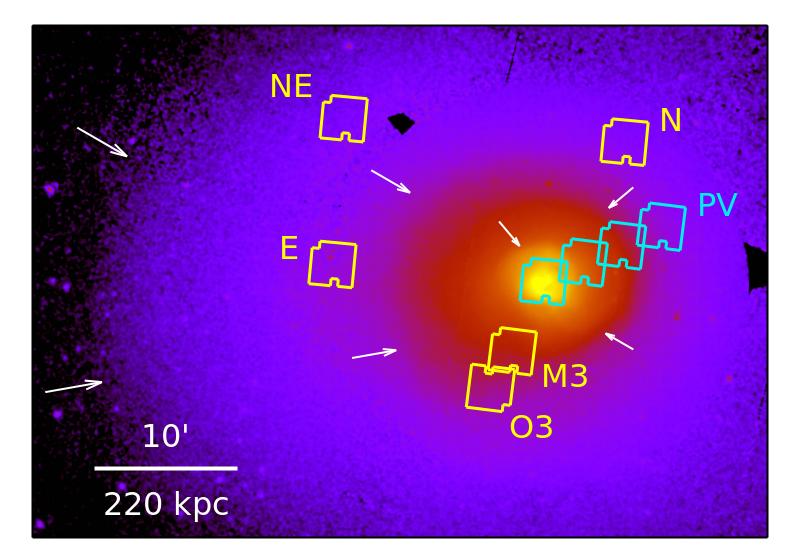}
\vspace{-5pt}
\caption{XRISM/Resolve pointings used in this work, labeled with their region names. The background shows the \xmm{} X-ray surface brightness in the $0.5-3.5\keV$ band \citep{Churazov2025}. White arrows indicate the sloshing cold fronts identified in the system \citep[e.g.,][]{Walker2018}. The pointings E, NE, and N (calibration), together with M3 and O3 (GO cycle 1 observations), are newly reported in this work, marked in yellow (see Section~\ref{sec:obs}). }
\label{fig:perseus}
\end{figure}

As one of the XRISM's Performance Verification (PV) targets, Perseus was observed along an NW arm of four pointings in February 2024 (see Fig.~\ref{fig:perseus}; \citealt{XRISM2025_Perseus}). The gas velocity radial profiles were measured up to the radius of $\simeq250\kpc$. They enable to unambiguously disentangle at least two dominant gas kinematic drivers operating on distinct spatial scales \citep{XRISM2025_Perseus}. The small-scale driver, confined within the central $\lesssim60\kpc$ of the cluster, is associated with AGN feedback; while the large-scale one (likely a few hundred kpc or even larger) dominates the outer region, driven by mergers of clusters. \citet{XRISM2025_Perseus} provided a lower limit of the energy injection scale, $\simeq100\kpc$, of the outer driver. A tighter constraint requires to map velocities over a broader region.

This paper reports five additional XRISM/Resolve pointings of the Perseus cluster observed in 2025 (see Fig.~\ref{fig:perseus} as well as Table~\ref{tab:params}). Combined with the PV data (observed in 2024), we present an extended, up-to-date velocity map. Perseus is currently the only XRISM cluster target with extensive mapping from its core region out to larger radii (a few hundred kpc), while still offering sufficient spatial resolution to resolve gaseous substructures (c.f., A2029, see \citealt{XRISM2025_A2029_Outer}). The coverage across a broad range of dynamical scales (i.e., $\simeq50-500\kpc$) and radial directions (E/NE, NW, and W) enables a robust characterization of the velocity field, shedding lights on physical properties of the dominant gas kinematic sources.

In this work, we focus on large-scale, merger-associated motions and show that their statistical features, e.g., velocity structure function, turbulent heating rate, can be effectively described by a velocity field dominated by a single, large-scale kinematic driver. The velocity measurements confirm the significant role of the turbulent cascade in merger energy conversion (from the gravitational to thermal energies). Our velocity maps also provide strong constraints on the merger scenario and viewing angles of the system, complementing other X-ray and weak-lensing measurements \citep{Churazov2025,HyeongHan2025}. This study highlights the scientific value of mapping the ICM and lays the groundwork for future projects and missions (e.g., XRISM key/legecy projects, HUBS, LEM, NewAthena; see \citealt{Cui2020,Kraft2022,Cruise2025}) aimed at mapping gas kinematics in nearby clusters with long exposures.

This paper is organized as follows. Section~\ref{sec:obs} outlines the XRISM/Resolve observations and data reduction. Section~\ref{sec:map} presents the extended velocity maps of Perseus. Section~\ref{sec:velocity} characterizes the velocity fields and their physical properties. Section~\ref{sec:merger} constrains the merger configuration of the cluster using the velocity map. Finally, Section~\ref{sec:conclusion} summarizes our main conclusions. Throughout the paper, we assume a flat cosmology with $H_0=67\kms\Mpc^{-1}$ and $\Omega_{\rm m}=0.32$, in which $1' = 22\kpc$ at the cluster redshift ($z\simeq0.017$).

\begin{table*}
\centering
\renewcommand{\arraystretch}{1.5}
\begin{minipage}{\linewidth}
\centering
\caption{Observed XRISM/Resolve calibration and GO pointings, their basic information (see also Table~\ref{tab:params_appendix}), and best-fit single-temperature ICM properties including $1\sigma$ statistical uncertainties. }
\label{tab:params}
\begin{tabular}{cccccccccc}
  \hline
  Name & $t_{\rm exp}$ (ks)\footnote{The net exposure.}
       & $R$ (kpc)\footnote{The projected radius of the pointing center. The E+NE region uses the geometric midpoint of the E and NE pointings as its center.}
       & $\ell_{\rm eff}$ (kpc)\footnote{The range of effective length scale, including the uncertainties characterized by the emissivity fraction range of $40-60\%$ (see Section~\ref{sec:velocity:qheat} and Appendix~\ref{sec:appendix:obs:leff}). }
       & $k_{\rm B}T$ (keV)\footnote{The ICM temperature.}
       & $Z_{\rm Fe}$\footnote{The Fe abundance relative to the \texttt{lpgs} proto-solar abundances \citep{Lodders2009}, see Fig.~\ref{fig:zfe}. }
       & $z$ ($\times10^{-2}$)\footnote{The ICM redshift.}
       & $u_{\rm los}$ ($\rm km\s^{-1}$)\footnote{The bulk velocity of the ICM in the rest frame of the central ICM ($z_{\rm icm}=0.017628$) with the heliocentric correction.}
       & $\sigma_{\rm los}$ ($\rm km\s^{-1}$)\footnote{The ICM velocity dispersion. For reference, the sound speed of $5-6\keV$ gas is $\simeq1100-1250\kms$.}
       & $f_{\rm nth}$ ($\times10^{-2}$)\footnote{The non-thermal pressure fraction (see Section~\ref{sec:velocity:fnth}).} \\
  \hline
  E   & 45 & 328 & $250-470$ & $5.54_{-0.30}^{+0.35}$ & $0.38_{-0.06}^{+0.07}$ & $1.862_{-0.025}^{+0.024}$ & $270_{-76}^{+72}$ & $330_{-67}^{+88}$ & $11.2^{+5.4}_{-3.7}$  \\
  NE  & 43 & 399 & $280-530$ & $5.35_{-0.41}^{+0.41}$ & $0.32_{-0.09}^{+0.10}$ & $1.890_{-0.061}^{+0.066}$ & $355_{-182}^{+199}$ & $660_{-392}^{+185}$ & $33.4^{+12.0}_{-24.5}$ \\
  E+NE\footnote{A combined fit of the E and NE pointings.}
      & 88 & 347 & $250-530$ & $5.47_{-0.23}^{+0.26}$ & $0.37_{-0.05}^{+0.06}$ & $1.866_{-0.027}^{+0.021}$ & $282_{-80}^{+62}$ & $349_{-57}^{+115}$ & $12.5^{+7.1}_{-3.4}$ \\
  N   & 50 & 243 & $190-400$ & $6.84_{-0.46}^{+0.43}$ & $0.35_{-0.06}^{+0.07}$ & $1.779_{-0.020}^{+0.012}$ & $22_{-61}^{+35}$ & $150_{-94}^{+39}$ & $2.0^{+1.2}_{-1.6}$   \\
  M3  & 122 & 112 & $70-260$ & $6.18_{-0.11}^{+0.12}$ & $0.44_{-0.02}^{+0.02}$ & $1.742_{-0.004}^{+0.003}$ & $-90_{-12}^{+10}$ & $171_{-12}^{+11}$ & $2.9^{+0.4}_{-0.4}$  \\
  O3  & 92 & 180 & $140-330$ & $6.32_{-0.20}^{+0.20}$ & $0.35_{-0.03}^{+0.03}$ & $1.786_{-0.011}^{+0.010}$ & $95_{-34}^{+30}$ & $275_{-30}^{+26}$ & $7.1^{+1.2}_{-1.3}$ \\
\hline
\hline
\vspace{-10pt}
\end{tabular}
\end{minipage}
\end{table*}

\section{XRISM/Resolve observations and data analysis} \label{sec:obs}

The Perseus cluster was observed multiple times during January/Feburary and July 2025 for calibration purposes and general observer (GO) project (Cycle 1, PI: I.~Zhuravleva; see Fig.~\ref{fig:perseus} and Table~\ref{tab:params_appendix}). Our data reduction and modeling largely follow the descriptions in \citet[][]{XRISM2025_Perseus} but using the latest versions of the HEASoft (v6.35.2), CalDB (v20250315), and AtomDB (v3.1.3).  The analysis was confined to the highest-resolution primary (Hp) events, providing spectral resolution of $\simeq4.5\eV$ (FWHM). The pixel 27 was excluded due to its gain drift issue \citep{Porter2024}.

Our spectra were extracted from the sub-FOV for four PV pointings or from the full FOV for calibration and GO pointings. Both source and background spectra were grouped to ensure at least one count per energy bin. For each region, we generated the corresponding L-size redistribution matrix file (RMF) and auxiliary response files (ARFs). We used an \xmm{} image in the ARF generation for the calibration/GO pointings, as some of them are not covered by \chandra{} data.

The grouped spectra were then fitted in the $3-11\keV$ energy range using a model that included the ICM, non-X-ray background (NXB), and, if necessary, AGN components. The ICM emission was modeled as a single-component spectrum from optically-thin plasma in collisional ionization equilibrium (\texttt{bvapec} model in \texttt{Xspec} v12.15.0), absorbed by the foreground along the line-of-sight (LOS) with a fixed equivalent hydrogen column density $n_{\rm H}=1.38\times10^{-21}\cm^{-2}$ \citep{Willingale2013}. The Fe~He$\alpha$ resonance ($\omega$) line was excluded from the \texttt{bvapec} (see Appendix~\ref{sec:appendix:obs:lremover}) and modeled separately with a redshifted Gaussian. Its redshift and width were both linked to the ICM component for all regions outside $60\kpc$.

\begin{figure}
\centering
\includegraphics[width=0.95\linewidth]{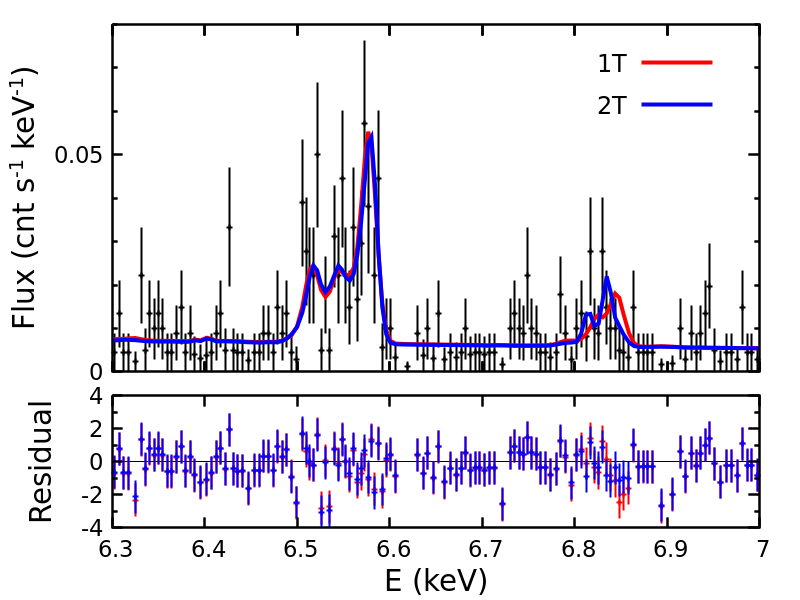}
\vspace{-5pt}
\caption{Fe~He$\alpha$ and Ly$\alpha$ lines in the E region with best-fit 1T and 2T models. The spectrum suggests the presence of an additional, more redshifted, hotter component, which requires confirmation by deeper observations. Residuals normalized by the statistical errors, i.e., (data-model)/error, are displayed in the lower panel (see Section~\ref{sec:obs} and Appendix~\ref{sec:appendix:obs:e2comp}).}
\label{fig:e2comp}
\end{figure}

\begin{figure*}
\centering
\includegraphics[width=0.9\linewidth]{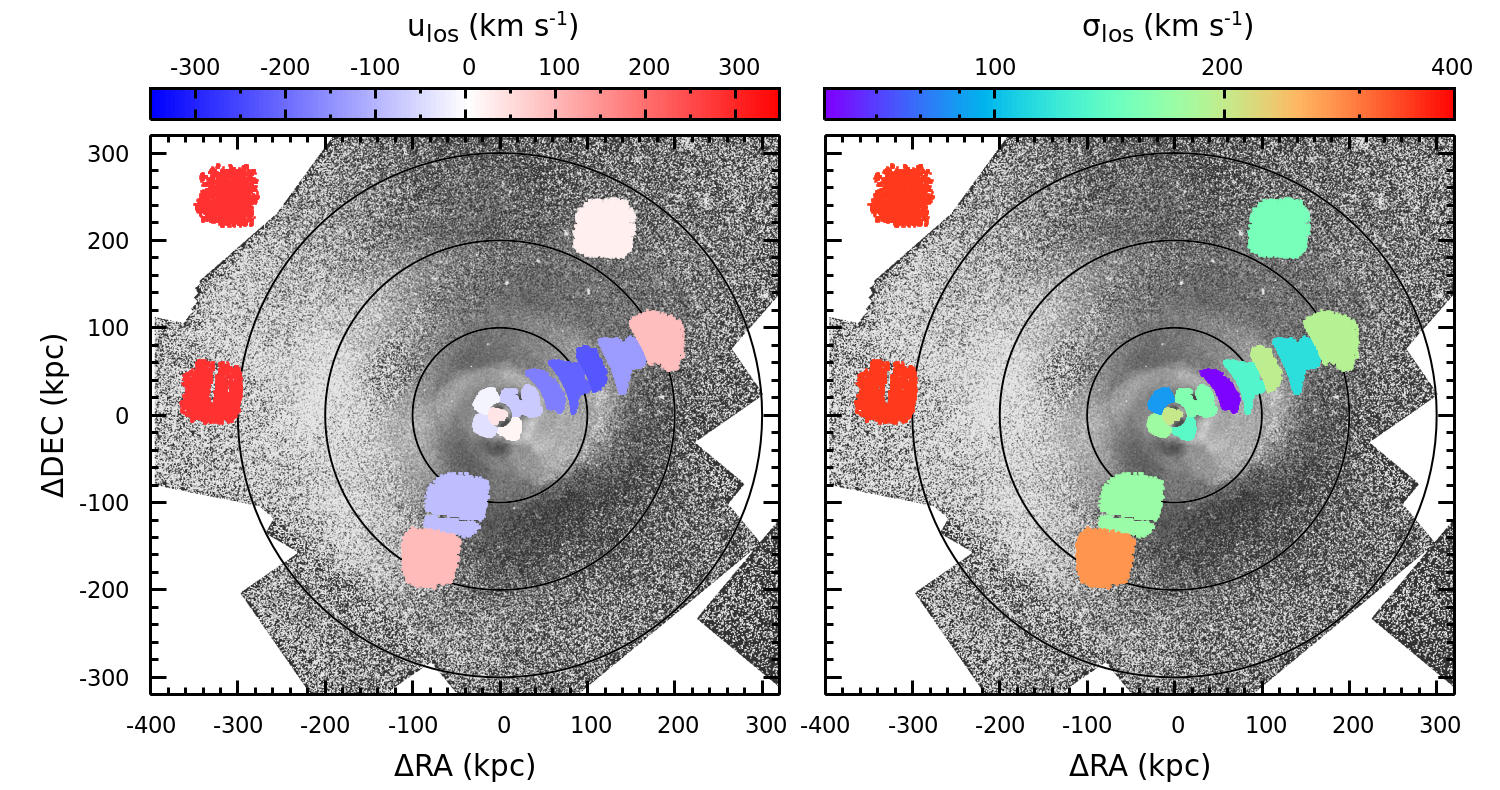}
\vspace{-5pt}
\caption{The best-fit gas bulk velocity (left) and velocity dispersion (right) maps of the Perseus cluster from nine XRISM/Resolve pointings. The maps are centered on the Perseus center (RA=49.9507, DEC=41.5117). The N and NE regions show the E+NE joint fitting result assuming a single temperature model. The color patches indicate the sky areas that contribute $50\%$ of the photons for the corresponding sub-regions based on raytracing simulations. The measurement uncertainties are shown in Fig.~\ref{fig:contours}. The \chandra{} X-ray residual is overlaid in the background highlighting the inner sloshing spiral and the eastern X-ray surface brightness excess, with black circles marking radii of 100, 200, and $300\kpc$. The bulk velocity distribution, in the rest frame of the central ICM ($z_{\rm icm}=0.017628$), shows an (asymmetric) dipole-like pattern, revealing a rotational motion of the ICM (see Section~\ref{fig:vmap}). }
\label{fig:vmap}
\end{figure*}

For the PV data, we adopted the same spatial binning scheme as \citet[][see their Extended Data Fig.~3]{XRISM2025_Perseus}.  A correction for the point spread function (PSF) effect (a.k.a., spatial-spectral-mixing, or SSM) was fully implemented (same as in \citealt{XRISM2025_Perseus}), which is particularly crucial for the radially-connected regions with steep X-ray surface brightness gradients (e.g., central regions of the cool-core clusters). We confirm that our new fitting results are fully consistent with \citet{XRISM2025_Perseus} within the $1\sigma$ uncertainty, despite the updated calibration and atomic databases.

For the calibration and GO data (newly reported in this work), we omitted the SSM correction due to the widely scattered sky positions of their pointings and tied all abundances (except He) to Fe as the emission lines from the non-Fe elements are relatively faint. Since these pointings are all located at least $5'$ away from the bright cluster center, they are much less affected by the SSM compared to the inner PV pointings, which is further confirmed with the Resolve raytracing simulations (see Section~\ref{sec:map}). We have examined the SSM correction between M3 and O3. Its effect on velocity measurements are smaller than $1\sigma$ statistic uncertainty. The best-fit ICM parameters are summarized in Table~\ref{tab:params}. The outermost pointings, E and NE, have relatively shallow exposures but exhibit consistent velocities. We therefore combined them (denoted as E+NE) to measure the averaged velocities near the E/NE regions (see Fig.~\ref{fig:contours_appendix} for a comparison between the joint and separate fittings). Unless stated otherwise, we use only this joint measurement for E/NE throughout the rest of the paper. Our conclusions remain largely unaffected if instead the NE region is excluded from the analysis.

We specifically examined the regions with high velocity dispersion for evidence of multiple temperature/velocity components. The region E shows a tentative indication of such complexity -- a $\sim2-3\sigma$ deviation near the Fe~Ly$\alpha$ lines, suggesting an additional, more redshifted, hotter component. A similar trend is seen in the region NE, but the data is considerably noisier. No significant sign of multiple components is found in other regions. In Fig.~\ref{fig:e2comp}, we compare the best-fit models with one-temperature (1T) and two-temperature (2T) components for the region E. A combination of cold ($k_{\rm B}T\simeq4.0\pm0.4\keV$) and hot ($k_{\rm B}T\gtrsim9\keV$) components slightly improve the fit of the Fe~Ly$\alpha$ lines (see Appendix~\ref{sec:appendix:obs:e2comp} for more details). The cold component dominates the Fe~He$\alpha$ line flux with $\sigma_{\rm los}\simeq290\pm70\kms$ and $u_{\rm los}\simeq200\pm70\kms$, consistent with our 1T result within $\simeq2\sigma$ (see the pink cross in Fig.~\ref{fig:contours}), while only an upper limit is obtained for the hot component's dispersion ($\simeq250\kms$) and a lower limit for the temperature. Deeper observations are required to confirm the result and to better separate the components if present. In this study, we focus on the 1T measurements for all regions unless noted otherwise, but will discuss how the potential temperature complexity in E/NE may influence our conclusions.

\section{Gas kinematic maps of Perseus} \label{sec:map}

\begin{figure}
\centering
\includegraphics[width=0.95\linewidth]{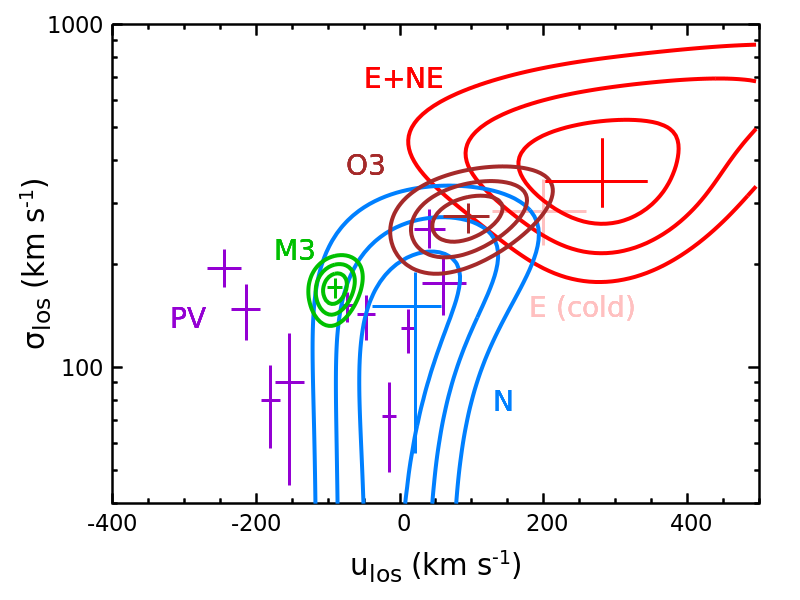}
\vspace{-5pt}
\caption{2D confidence contours (1, 2, and 3$\sigma$) of the bulk velocity and velocity dispersion measurements of our new regions (calibration and GO pointings). The red contours indicate a joint fit between the E and NE regions (see Fig.~\ref{fig:contours_appendix} for their individual fits). The crosses mark the best-fit parameters and their $1\sigma$ uncertainties, including also all PV data (ten sub-regions; see Section~\ref{sec:map}). The pink cross indicates the cold ICM component in the 2T model for region E (see Section~\ref{sec:obs}).}
\label{fig:contours}
\end{figure}

Fig.~\ref{fig:vmap} shows our extended velocity maps of the hot ICM in Perseus, assuming a single-temperature component, based on four PV, three calibration, and two GO pointings.\footnote{As of the preparation of this paper, the C3 pointing has also been completed. An in-depth joint analysis between the full NW and S arms will be presented in a companion paper, featuring gas velocities within the Perseus cool core.}  We used the built-in HEASoft tool \texttt{xmatraceback} to trace the sky region contributing scattered photons to each detector region where we extracted spectra. In Fig.~\ref{fig:vmap}, the color patches indicate the sky regions that contribute $50\%$ of the photons for their corresponding detector regions, reflecting approximately the areas where we measured the velocities. The confidence levels of our velocity measurements are presented in Fig.~\ref{fig:contours}. The pink cross shows the cold component in the 2T model of the region E for comparison.

The left panel of Fig.~\ref{fig:vmap} shows the LOS bulk velocity with heliocentric correction in the rest frame of the central ICM, $z_{\rm ICM}=0.017628$, measured within $\simeq30\kpc$ (the central pointing; see \citealt{XRISM2025_Perseus}). Such rest frame is selected to highlight gaseous substructures in the Perseus core, which does not affect any conclusions in the paper. Note that the redshift of the brightest cluster galaxy (BCG) stellar component is $z_{\rm BCG}=0.017284$ \citep{Hitomi2018_Velocity}, corresponding to $\simeq-103\kms$ velocity offset relative to the central ICM. An asymmetric dipole-like structure is visible in the bulk velocity distribution, where the most positive (moving away from us) and negative (towards us) velocities are located on the opposite sides of a symmetry axis oriented near the north-south direction. It implies a gas bulk rotation on a scale of at least $\sim400\kpc$, associated with the recent merger processes. Interestingly, the circumnuclear disk in Perseus is rotating along a similar direction on much smaller scales ($<0.1\kpc$; see \citealt{Nagai2019}). In Section~\ref{sec:merger}, we use this dipole-like feature to constrain sloshing configuration and viewing angles of the system. \citet{Sanders2020} mapped the bulk velocity of Perseus over a $\sim600\kpc$ radial range with \xmm{} observations by calibrating the energy scale of the EPIC-pn detector, but the statistical uncertainties are relatively large -- typically $\gtrsim200\kms$ ($1\sigma$; see their fig.~16) and $\gtrsim300\kms$ near our E, NE, and N regions.

The right panel of Fig.~\ref{fig:vmap} shows the velocity dispersion map. The region E+NE exhibits the highest velocity dispersion detected so far in Perseus (with $>3\sigma$ significance), suggesting strong gas random motions and high nonthermal pressure (see Section~\ref{sec:velocity:fnth}) in its vicinity. Coherent bulk motions, especially overlapping velocity shears near sloshing cold fronts or merger shocks, can contribute to the velocity dispersion. However, there are no clear signs of such structures from high-resolution X-ray images, aside from an extended X-ray surface brightness excess in the nearby regions (see Appendix~\ref{sec:appendix:obs:e2comp} for a discussion on multiple temperature components). Interestingly, the O3 region, partially overlapping with the eastern excess, shows also high velocity dispersion, which hints the chaotic behavior of the X-ray excess and its connection with the recent mergers.

\section{Characterizing the kinematic properties} \label{sec:velocity}

Gas velocity field offers valuable insights into the assembly history of the Perseus cluster and its ICM properties. In this section, we characterize the statistical properties of our velocity measurements and discuss their physical implications. We caution that the statistical significance of our findings remains limited by the current spatially sparse velocity sampling. Additional XRISM observations with a more uniform mapping configuration will be essential for confirmation.

\subsection{Nonthermal pressure} \label{sec:velocity:fnth}

\begin{figure}
\centering
\includegraphics[width=0.95\linewidth]{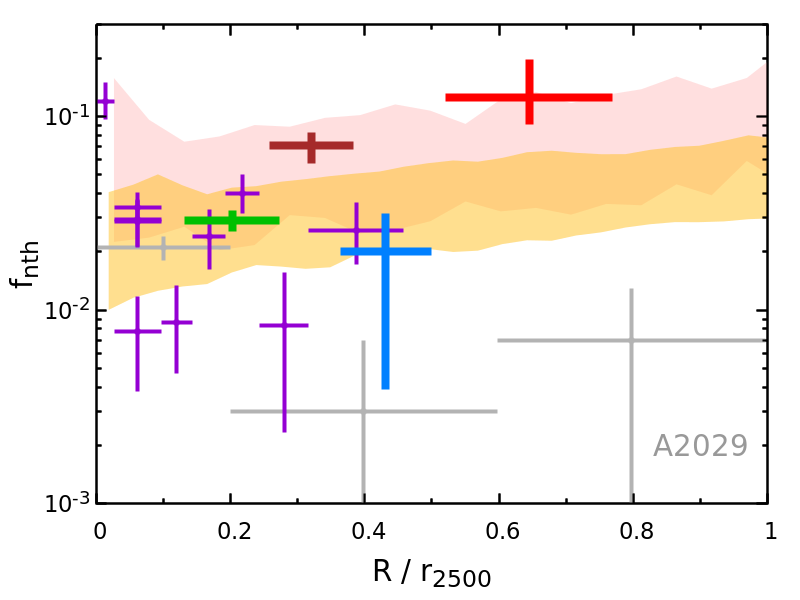}
\vspace{-5pt}
\caption{Nonthermal pressure fraction profile of the Perseus cluster, compared with A2029 (grey points; \citealt{XRISM2025_A2029_Outer}). Colors follow the same scheme as in Fig.~\ref{fig:contours} for Perseus. The shaded bands show numerical predictions from TNG cosmological simulations for two cluster subsamples: (i) the 31 most relaxed clusters from the TNG-300 simulation suite (yellow) and (ii) 30 Perseus-like massive, cool-core clusters from TNG-Cluster suite (pink). The bands represent the 10th-90th percentile range of the $f_{\rm nth}$ distribution within each sample. The inner AGN-dominant region, as well as the regions overlapping the eastern X-ray surface brightness excess (E+NE and O3), show high $f_{\rm nth}$ ($>7\%$; see Section~\ref{sec:velocity:fnth}). }
\label{fig:nonthermal}
\end{figure}

Nonthermal pressure fraction ($f_{\rm nth}$), the ratio of kinematic to total pressure in the ICM, characterizes the dynamical significance of gas random motions and the extent to which the cluster atmosphere deviates from hydrostatic equilibrium \citep[e.g.,][]{Evrard1990,Nagai2007,Lau2009,Shi2014}. Assuming isotropic velocity fields and the LOS velocity dispersion predominantly reflects gas random motions, the nonthermal pressure fraction solely depends on the 1D Mach number $M_{\rm 1D}\,(\equiv\sigma_{\rm los}/c_{\rm s})$ as
\be
f_{\rm nth}=\frac{\mathcal{M}_{\rm 1D}^2}{\mathcal{M}_{\rm 1D}^2 + \gamma^{-1}},
\ee
where $c_{\rm s}\,(\equiv\sqrt{\gamma k_{\rm B}T/\mu m_{\rm p}})$ is sound speed; $\gamma\,(=5/3)$, $\mu\,(=0.6)$, and $m_{\rm p}$ are adiabatic index of monatomic gas, mean molecular weight per ion, and proton mass, respectively.

Most XRISM cluster studies so far suggest low nonthermal pressure contributions in the cluster cores, e.g., $\simeq1-4\%$ along the LOS across the center of the Coma Cluster, Centaurus, A2029, and Ophiuchus \citep{XRISM2025_Coma,XRISM2025_Centaurus,XRISM2025_A2029_Inner,Fujita2025}. Interestingly, deep off-center observations of A2029 reveals a decreasing nonthermal pressure fraction with the radius up to at least $r_{2500}$ along the NE direction,\footnote{The $r_{\Delta}$ represents the radius within which an average mass density is $\Delta$ (e.g. 2500, 500, 200) times the critical density of the Universe at the cluster redshift.} different from cosmological simulation predictions in general, despite that A2029 is known as one of the most relaxed clusters in observations.

Perseus is the second cluster for which we attempt to measure the nonthermal pressure profile out to approximately $0.7r_{2500}$ ($\simeq570\kpc$; \citealt{deVries2023}) after A2029. The results are shown in Fig.~\ref{fig:nonthermal}, where $f_{\rm nth}$ ranges from $\simeq0.5-5\%$, comparable to values observed in most other XRISM clusters, except in the inner- and outermost regions ($\gtrsim10\%$) and in O3 ($\simeq7\%$).
The innermost region is strongly influenced by radio-mode AGN feedback (e.g., jets, bubbles). The four neighbouring regions within the annulus $R\simeq10-60\kpc$ show azimuthal variations by a factor of $\sim6$ (see Fig.~\ref{fig:vmap}), plausibly also associated with the central supermassive black hole (SMBH) activities (see more discussions in \citealt{XRISM2025_Perseus}). We note that there are potential systematic uncertainties induced by the imperfect PSF calibration, which tend to have a stronger impact on smaller detector regions. Although \citet{XRISM2025_Perseus} showed that $30\%$ off-axis ARF uncertainties affect velocity measurements still within their $1\sigma$ statistic errors (see their extended data fig.~4), the impact might depend on specific spatial regions and binning schemes.

The E+NE and O3 regions exhibit the highest nonthermal pressure fractions in the current Perseus measurements. In the 2T model of region E, the dominant cold component yields $f_{\rm nth}\simeq11\pm5\%$, consistent with the 1T fit. Perseus is often considered a relaxed system, yet it also shows active AGN feedback and prominent sloshing features in its X-ray distribution. For comparison, we show numerical predictions based on the IllustrisTNG cosmological simulations in Fig.~\ref{fig:nonthermal} \citep{Pillepich2018,Springel2018,Naiman2018,Marinacci2018,Nelson2018,Nelson2019}. They are estimated based on the X-ray-weighted velocity dispersion for a direct comparison with observations. Two subsamples of the TNG clusters are considered: one consists of the 31 most relaxed clusters from the TNG-300 simulation suite, selected based on the spherical gas morphology and absence of substructures, with masses ranging $1-7\times10^{14}\msun$ (yellow; see \citealt{Zhang2024} for details); the other consists of 30 Perseus-like clusters from the TNG-Cluster suite, selected based on halo mass ($5-10\times10^{14}\msun$) and the presence of cool cores (pink; see \citealt{Truong2024,XRISM2025_Simulation}). The shaded bands indicate the 10th and 90th percentiles of the $f_{\rm nth}$ scatter within each sample. The Perseus-like clusters from the TNG-Cluster exhibit systematically higher nonthermal pressure fraction than the relaxed cluster sample. Observationally, Perseus resembles the relaxed sample within $\simeq0.2r_{2500}$ (excluding the innermost region), with values markedly lower than those of its massive, cool-core numerical counterparts. The same holds for the entire NW arm. Beyond $0.2r_{2500}$, only the two regions overlapping with the eastern X-ray excess show higher $f_{\rm nth}$ than the relaxed sample, consistent with the TNG-Cluster one.

We further compare Perseus and A2029 -- the two clusters with the most extended nonthermal pressure measurements to date and with comparable effective length radial profiles (see Fig.~\ref{fig:leff}). The effective length characterizes the region size along the sightline that contributes majority of the X-ray flux (\citealt{Zhuravleva2012}; see also Appendix~\ref{sec:appendix:obs:leff}). The measured velocity dispersion depends on the effective length $\ell_{\rm eff}$ along the LOS. In observations, a fair comparison of $f_{\rm nth}$ is possible only when the clusters (or regions) being compared have similar $\ell_{\rm eff}$.

Perseus and A2029 share a collection of similar gaseous features shaped by AGN-driven bursts and mergers. They both exhibit prominent radio lobes (on $\sim10$ and $30\kpc$ scales, respectively) powered by their central SMBHs \citep[e.g.,][]{Timmerman2022}, as well as sloshing spirals within $\simeq200\kpc$ ($\sim0.3r_{2500}$). These similarities partially explain the comparable $f_{\rm nth}$ of the two clusters within the radial extent of their inner sloshing spirals. Beyond the spiral, however, the velocity dispersion in A2029 drops rapidly \citep{XRISM2025_A2029_Outer}, unlike in Perseus -- a difference reflects the more complex merger history of Perseus. Simulations suggest that a minor merger occurred $\sim2-3\Gyr$ ago in A2029,\footnote{A detailed numerical modeling study of A2029 will be presented in a separate paper.} whose sloshing spiral has not yet fully developed (see also \citealt{Sohn2019}). In Perseus, the inner and outer sloshing cold fronts may have been driven by separate mergers, with a $\sim3-5\Gyr$ time interval, resulting in a more turbulent atmosphere outside the cool core than in A2029 (see Section~\ref{sec:merger}).

\subsection{Velocity structure function and injection scales} \label{sec:velocity:vsf}

The velocity structure function is an essential diagnostic tool for characterizing gas motions utilizing the LOS bulk velocity map. It quantifies the velocity differences between spatially separated regions, thereby probing directly the scale dependence of turbulent flows \citep[e.g.,][]{Zhuravleva2012,Miniati2014,Zuhone2016,Zhang2024,Gatuzz2023}. This method has been applied to the recent XRISM observations of the Coma cluster and revealed tentatively a sign of extremely steep velocity power spectrum \citep{XRISM2025_Coma}, although whether it reflects cosmic variance and/or inhomogeneity in the velocity field is still under debate \citep{Vazza2025}.

The nine pointings of Perseus provide the most powerful dataset to date for a similar analysis (c.f., two pointings in Coma). Perseus, however, is a cool-core cluster, exhibiting strong X-ray surface brightness gradients that modulate the observed structure function. \citet{XRISM2025_Perseus} also suggested the presence of at least two drivers within the Perseus cool core. Therefore, in our estimation, we adopted only the region beyond $R\simeq60\kpc$ to avoid the impact of AGN feedback (one of the dominant drivers). To interpret the measurements, we generate Gaussian random velocity fields with underlying energy power spectra $E(k)$ as baseline models -- a similar approach has been applied in the study of the Coma cluster \citep{XRISM2025_Coma},
\be
E(k) = E_0 \frac{(k\ell_{\rm inj})^{\alpha}}{1+(k\ell_{\rm inj})^{\alpha-\beta}}e^{-k\ell_{\rm diss}},
\label{eq:Ek}
\ee
where $E_0$ is a normalization factor, $\ell_{\rm inj}$ is the energy injection scale, and $\ell_{\rm diss}$ is a sufficiently small dissipation scale ($\ll\ell_{\rm inj}$, fixed at $1\kpc$ in this study), $\alpha$ and $\beta$ are the spectral indices within the inertial range ($\alpha=-5/3$ if Kolmogorov turbulence, see \citealt{Kolmogorov1941}) and at large scales ($>\ell_{\rm inj}$), respectively. \citet{Subramanian2006} suggested $\beta\simeq2$ as the ICM is mildly compressible, but it might be sensitive to the detailed physics of the system (e.g., gravitational stratification). The cluster environment can be far more complicated than isotropic turbulence (see \citealt{Hosking2022} and references therein for a detailed discussion of the latter situation). We project the 3D Gaussian random velocity field along the LOS, assuming Perseus-like X-ray emissivity distribution. To examine the effect of gravitational stratification on turbulence evolution,  we further run ``turbulence-in-a-box'' simulations, similar to those performed in \citet{XRISM2025_Perseus}, and found little difference between the simulations and the Gaussian model regarding the shape of the velocity structure function (see Appendix~\ref{sec:appendix:sim} for more details).

\begin{figure}
\centering
\includegraphics[width=0.95\linewidth]{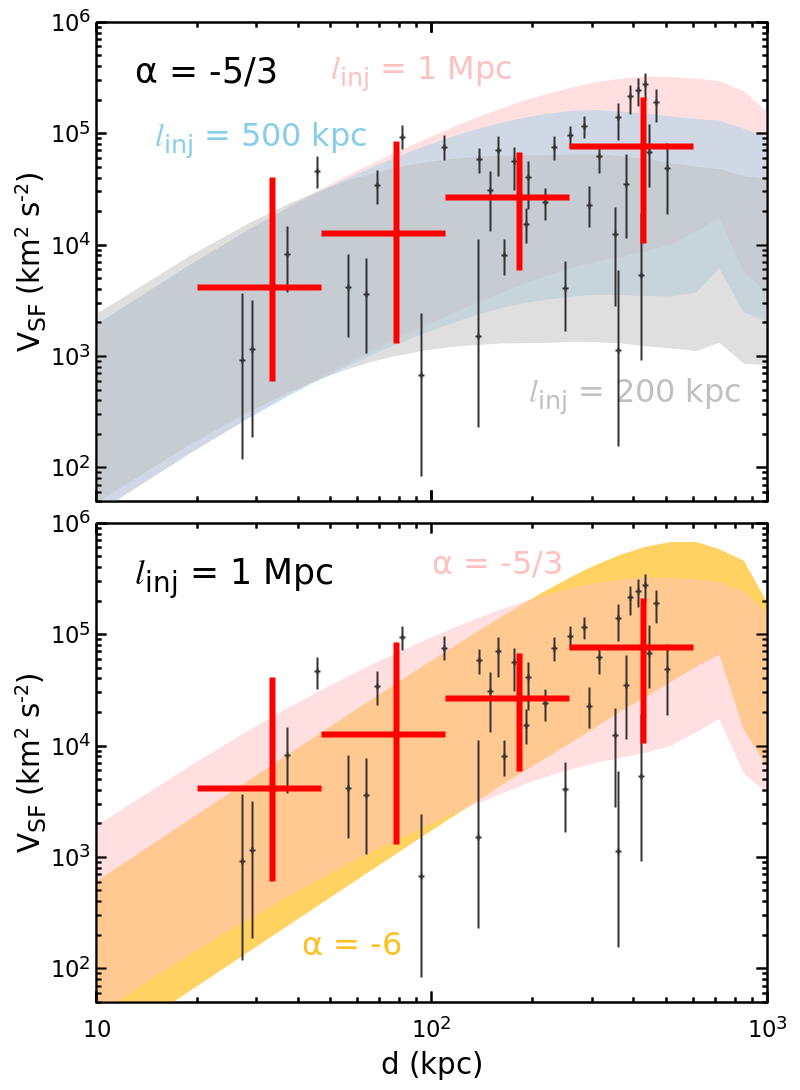}
\vspace{-5pt}
\caption{The second-order velocity structure function for the velocity field outside the AGN-dominant region ($\gtrsim60\kpc$) in Perseus, grouped into four bins (red points). Black points show all available velocity pairs with their $1\sigma$ statistical uncertainties. The shaded regions show theoretical predictions and their $68\%$ scatters based on Gaussian random fields. Various energy injection scales (top panel) and spectral indices are shown to compare with the observations. The XRISM Perseus dataset favors a large injection scale (at least a few hundred kpc) but is not sensitive to the spectral index (see Section~\ref{sec:velocity:vsf}). }
\label{fig:vsf}
\end{figure}

Fig.~\ref{fig:vsf} shows the second-order velocity structure function constructed from both observations and Gaussian models, i.e.,
\be
V_{\rm SF}(d) = \langle|u_{\rm los}(\textbf{R}+\textbf{d})-u_{\rm los}(\textbf{R})|^2\rangle,
\ee
where $\langle\rangle$ denotes an average over all velocity pairs on the sky with spatial separation $d\ (=|\textbf{d}|)$. The black points represent all available velocity pairs derived from our velocity map; the error bars correspond to $1\sigma$ statistical uncertainties in the measurements. To estimate the structure function, we grouped these velocity pairs into four separation bins, shown as the red points. We also tested three and five bins, which yielded similar results. To evaluate uncertainties, we performed Monte Carlo sampling of all velocity differences within each bin. The presented error bars enclose the $68\%$ scatter range, accounting for both variations in the velocity differences and their measurement errors. Our theoretical predictions are overlaid as shaded bands, indicating $68\%$ scatters that reflect the cosmic variance based on 200 random realizations. We experimented various energy injection scales (top panel) and spectral indices (bottom panel). We did not fit the velocity normalization but adjusted it to visually match the measurements. Our models were derived from the X-ray-weighted bulk velocity within $R=60-400\kpc$, a radial range consistent with the observations. We also tested estimating $V_{\rm SF}$ from regions similar to those in the observations, which resulted in generally consistent structure functions. However, we emphasize that, in both cases, our models contain far more velocity pairs than the observations. The current data may therefore be affected by cosmic variance.

A single, large-scale driver can effectively reproduce the observations. It corresponds to the outer, merger-associated driver proposed by \citet{XRISM2025_Perseus} in their ``double-cascade scenario'', which can be a combination of multiple drivers as long as they all operate at large scales. Our measurements favor a large injection scale ($\ell_{\rm inj}\sim500\kpc$ or even larger), broadly consistent with cosmological simulation predictions \citep[e.g.,][]{Norman1999,Vazza2009,Miniati2014,Zhang2024}. The bin with the largest separation provides the strongest constraint, with 7 out of 15 velocity pairs lying above the $1\sigma$ limit ($84\%$) of the $\ell_{\rm inj} = 200\kpc$ prediction. Assuming the pair distribution is Gaussian and independent, our measurements can rule out $\ell_{\rm inj} = 200\kpc$ at a confidence level of $\gtrsim 99.5\%$. Although the model of $\ell_{\rm inj} = 200\kpc$ can be shifted vertically to match data better at large separations, it will lead to an unrealistic velocity amplitude. Note that regions E/NE contribute dominantly to the bin with $d\gtrsim300\kpc$. If the presence of multiple temperature components is confirmed, additional complexity arises. Excluding E/NE from our velocity structure function estimation, we cannot rule out $\ell_{\rm inj} = 200\kpc$ based on the remaining dataset. \citet{XRISM2025_Perseus} used the velocity dispersion dip near $R\simeq60-100\kpc$ along the NW direction to place a lower limit of the $\ell_{\rm inj}$ ($\gtrsim100\kpc$). Their argument is that the $\ell_{\rm inj}$ cannot be smaller than the local effective length scale, otherwise the radial profile of the velocity dispersion would be flat. We confirm this result with the velocity structure function independently, and our new data slightly improve their constraint. Since information at larger separations (e.g., $1\Mpc$) is critical to examine $\ell_{\rm inj}=200-500\kpc$, future observations to the west/southwest of the cluster will provide key additional insights.

We have also tested a very steep spectrum inspired by the recent Coma measurements \citep{XRISM2025_Coma} as a possibility of a large dissipation scale ($\gtrsim240\kpc$). However, our measurements cannot distinguish it from the Kolmogorov scaling due to the insufficient number of velocity pairs with small separations. In \citet{XRISM2025_Coma}, bulk velocity and velocity dispersion were combined to constrain the velocity power spectrum. We defer such an analysis to future work owing to the complex surface brightness stratification in Perseus.

\subsection{Turbulent heating rate and dissipation energy} \label{sec:velocity:qheat}

\begin{figure}
\centering
\includegraphics[width=0.95\linewidth]{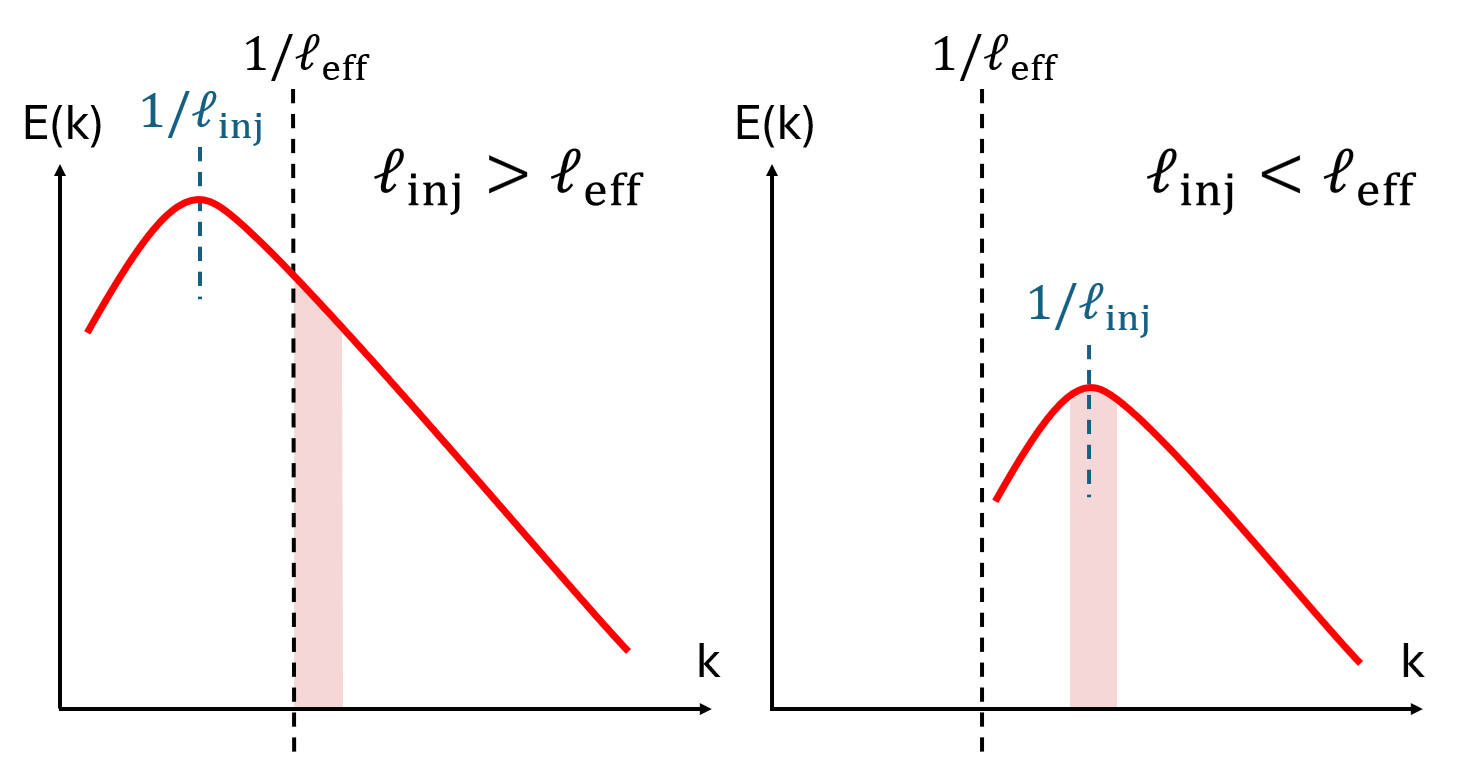}
\vspace{-5pt}
\caption{A sketch illustrating the effect of X-ray emissivity distribution on modulating the dominant velocity power spectrum we probe. The X-ray emissivity acts as a velocity high-pass filter, allowing us to probe only the largest scales smaller than $1/\ell_{\rm eff}$ in $k$-space (see Appendix~\ref{sec:appendix:obs:leff} for definition and calculation of $\ell_{\rm eff}$). The pink regions depict the range of spatial scales that dominate the gas kinematic energy in the measurements. In this work, we focus on merger-driven, large-scale gas motions, which likely correspond to the situation shown on the left (see Section~\ref{sec:velocity:qheat}). }
\label{fig:sketch_leff}
\end{figure}

Given the suggested large energy injection scales, we estimated the turbulent heating rate following \citet{XRISM2025_Perseus}, i.e.,
\be
Q_{\rm heat}=C_0\frac{\rho_{\rm gas}\sigma_{\rm los}^3}{\ell_{\rm eff}},
\label{eq:heating_rate}
\ee
where $C_0$ is a constant coefficient, $\rho_{\rm gas}$ is the gas mass density averaged along the LOS, and $\ell_{\rm eff}$ is the effective length (see Appendix~\ref{sec:appendix:obs:leff}). The X-ray emissivity distribution plays a role as a velocity high-pass filter, so that the measured X-ray-weighted $\sigma_{\rm los}$ reflects gas motions on the scale of $\ell_{\rm eff}$ or smaller. In a turbulent cascade scenario, gas kinetic energy within the atmosphere is dominated by motions on the largest scales (see a sketch in Fig.~\ref{fig:sketch_leff}). Eq.~\ref{eq:heating_rate} is valid for motions with $\ell_{\rm inj}\gtrsim\ell_{\rm eff}$, which is likely the situation for mergers. The denominator in Eq.~\ref{eq:heating_rate}, however, needs to be replaced by the turbulent energy injection scale $\ell_{\rm inj}$ when $\ell_{\rm inj}<\ell_{\rm eff}$, a possible situation in the AGN-feedback-dominant region (see more discussions in \citealt{XRISM2025_Perseus}). We adopted $C_0=5$, calibrated by laboratory experiments and direct numerical simulations of hydrodynamic turbulence (see \citealt{Zhuravleva2014} and references therein). Its exact value is not critical for our order-of-magnitude estimations, as long as it remains within an order of unity.

\begin{figure}
\centering
\includegraphics[width=0.95\linewidth]{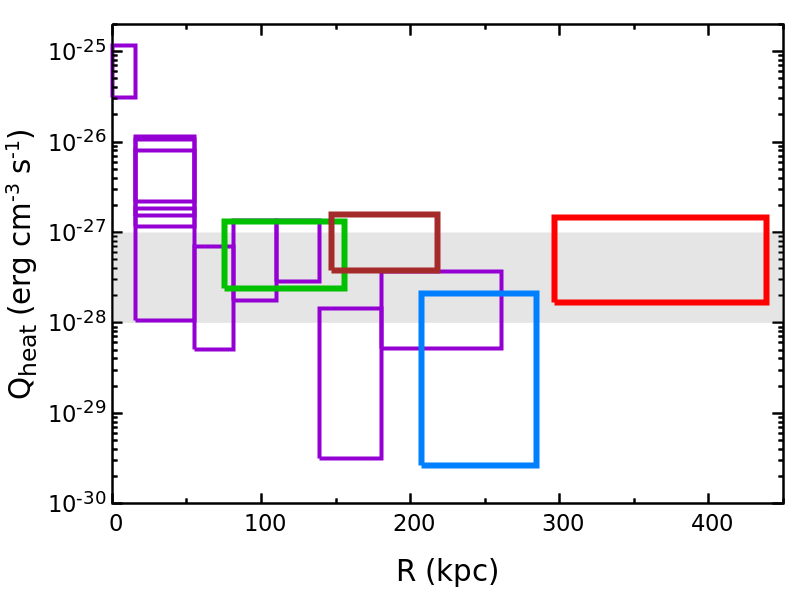}
\vspace{-5pt}
\caption{Radial profile of the turbulent heating rate. The color scheme is the same as in Fig.~\ref{fig:contours}. The uncertainties incorporate both the velocity statistical errors and effective length variations within each region. Outside the AGN-dominant region (i.e., $\gtrsim60\kpc$), the heating rate appears approximately uniform ($\sim10^{-28}-10^{-27}\erg\cm^{-3}\s^{-1}$, marked by the grey band), covering a dynamical range of $\ell_{\rm eff}\simeq100-400\kpc$ (see Section~\ref{sec:velocity:qheat}). }
\label{fig:heating_rate}
\end{figure}

The estimated heating-rate radial profile is shown in Fig.~\ref{fig:heating_rate}. The error bars include both the statistical uncertainties ($1\sigma$) of the velocity dispersion measurement and the spatial variations of the effective length (and their uncertainties) within each region. We assume the dominant injection scale is larger than or comparable to the maximum $\ell_{\rm eff}$ within our radial coverage ($\sim 500\kpc$; see Table~\ref{tab:params}), which is supported by the velocity structure function discussed in Section~\ref{sec:velocity:vsf}.
The radial profile of the heating rate is strongly peaked at the cluster center and flattens outside $R\simeq60\kpc$, forming a plateau that extends to the maximum radius we probe (see \citealt{XRISM2025_Perseus} for more discussions on the central peak associated with the AGN heating). Within the still-large uncertainties, all regions in the plateau are consistent with a uniform heating rate of approximately $10^{-28}-10^{-27}\erg\cm^{-3}\s^{-1}$ (estimated visually), though the regions near the far end of the NW arm appear lower. A defining feature of Kolmogorov-like turbulence with a single injection scale is the assumption of a constant heating rate within the inertial range \citep{Kolmogorov1941}. Our measurement does not significantly deviate from this scenario. This consistency is generally in line with the fact that most of the kinetic energy is released during the early stage of a merger, e.g., within the first dynamical timescale. The turbulence decays over a much longer timescale. Both $\ell_{\rm inj}$ and its corresponding velocity amplitude ($\sigma_{\rm inj}$) evolve with time and the eddy turnover timescale $\Delta t_{\rm turnover}\,(=\ell_{\rm inj}/\sigma_{\rm inj})$ increases as $\sim t$, independent of the spectral slope on large scales (i.e., $\beta$ in Eq.~\ref{eq:Ek}, see \citealt{Subramanian2006}).

We can then characterize the contribution of turbulent dissipation in the merger energy budget. The turbulent dissipation energy can be approximated as
\be
E_{\rm diss} \simeq \ell_{\rm inj}^3\cdot  Q_{\rm heat}\cdot \Delta t_{\rm turnover}.
\label{eq:e_diss}
\ee
In the early stage of a merger, $\ell_{\rm inj}$ reflects either the size of the infalling subhalo or the characteristic scale of its orbit. Under an assumption of the Kolmogorov scaling, i.e., $\sigma_{\rm inj}=(\ell_{\rm inj}/\ell_{\rm eff})^{1/3}\sigma_{\rm los}$ within the inertial range, the equation is further expressed as
\begin{multline}
E_{\rm diss} \simeq 5\times10^{62}\erg\,\Big(\frac{\ell_{\rm inj}}{1\Mpc}\Big)^{11/3}\Big(\frac{\ell_{\rm eff}}{100\kpc}\Big)^{1/3} \\
\Big(\frac{ Q_{\rm heat}}{10^{-28}\erg\cm^{-3}\s^{-1}}\Big)\Big(\frac{\sigma_{\rm los}}{100\kms}\Big)^{-1}.
\end{multline}
If the energy power spectrum is close to the form of Eq.~\ref{eq:Ek}, $E_{\rm diss}$ scales as $t^{(4-2\beta)/(3+\beta)}$, remaining time-independent for $\beta=2$, in which case the measured $E_{\rm diss}$ is temporally representative \citep{Subramanian2006}.
We argue that, even if $E_{\rm diss}$ evolves with time, our estimate is likely close to an average over the entire merger period, since the $\sigma_{\rm los}$ implies a reasonable dissipation timescale, i.e., $\Delta t_{\rm turnover}\simeq2$, 4, 3, and $2\Gyr$ from the region E+NE, N, M3, and O3, assuming $\ell_{\rm inj}=1\Mpc$. For reference, the Perseus's dynamical timescale is $\sim r_{200}^{3/2}/\sqrt{GM_{200}}\simeq1.5\Gyr$, where $G$ is the gravitational constant, $M_{200}\ (\simeq6\times10^{14}\msun)$ and $r_{200}\ (\simeq1.8\Mpc)$ are the virial mass and virial radius of Perseus, respectively. A merging cluster generally requires $\sim3-4$ dynamical timescale to dissipate its kinetic energy and relax \citep[e.g.,][]{Poole2006}. As a comparison, the situation in A2029 is different. Its outer pointings correspond to $\Delta t_{\rm turnover}\simeq8\Gyr$, assuming the same $\ell_{\rm inj}$ as in Perseus, which is likely only a lower limit. This implies that the gas environment outside the sloshing spiral in A2029 has been largely settled after an earlier merger event.

The XRISM measurements suggest $E_{\rm diss}\sim10^{62}-10^{63}\erg$ in Perseus, an order-of-magnitude estimation. It can be compared with the gravitational energy released through a merger into the ICM of Perseus
\be
E_{\rm grav} \simeq f_{\rm gas}\frac{G\,M_{200}^2}{\xi\,R_{200}},
\ee
where $f_{\rm gas}\ (\simeq0.17)$ is the gas fraction of the cluster and $\xi$ is the merger mass ratio. A merger with $\xi\simeq3-10$ (as suggested in Section~\ref{sec:merger}, see also \citealt{Bellomi2024}) results in $E_{\rm grav}\simeq3\times10^{62}-10^{63}\erg$, comparable to $E_{\rm diss}$, implying the significant role of turbulent dissipation in the merger energy conversion. This is in line with theoretical expectations \citep[e.g.,][]{Vazza2011,Miniati2015,Shi2020}. XRISM/Resolve, for the first time, allows such an observational examination.

\begin{figure*}
\centering
\includegraphics[width=0.9\linewidth]{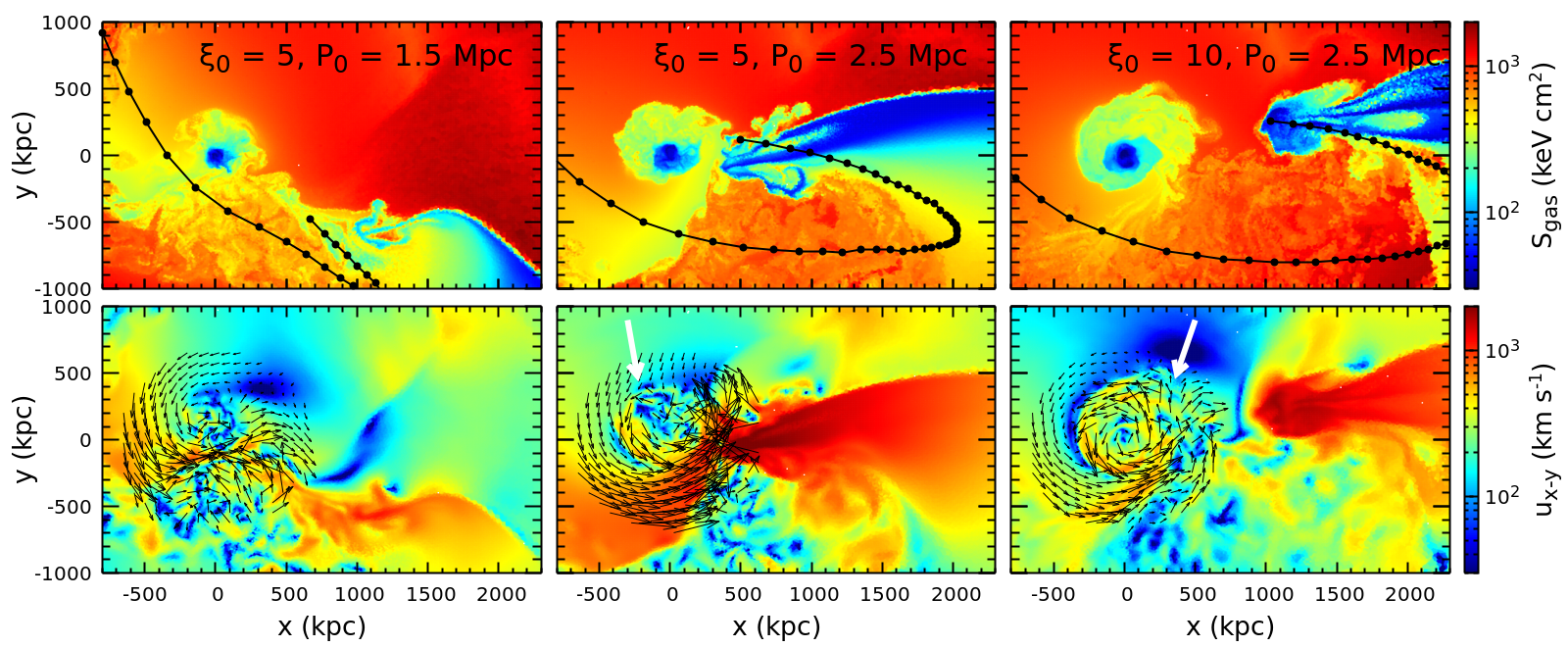}
\vspace{-5pt}
\caption{Slices of the gas entropy (top) and velocity in the $x-y$ plane (bottom) from the simulations $(\xi_0,\,P_0)=(5,\,1.5\Mpc)$, $(5,\,2.5\Mpc)$, and $(10,\,2.5\Mpc)$ from left to right at $t=3.1,\ 3.2,\ 5.0\Gyr$, respectively. All slices are taken through the merger plane (the $x-y$ plane). Subcluster trajectories in each simulation are shown by points connected with lines in the top panels, traced from snapshots at $0.1\Gyr$ intervals. The black vectors in the bottom panels illustrate both direction and amplitude of the velocity field in the merger plane (within $\simeq600\kpc$). The white arrows indicate the LOS directions (projected in the $x-y$ plane) adopted in Fig.~\ref{fig:merger_mock}. The merger with the smallest subhalo and largest impact parameter produces the most regular, least turbulent sloshing spirals in the cluster core (see Section~\ref{sec:merger:sim}). }
\label{fig:merger_zplane}
\end{figure*}

Finally, we note that the validity of our estimation depends on whether the observable, $\sigma_{\rm los}$, reflects the strength of the gas random motions. It is a reasonable assumption for the systems like Perseus. The most recent merger in Perseus occurred $3-5\Gyr$ ago (see Section~\ref{sec:merger}), allowing large-scale turbulence and its cascade to become well established. Newly formed large-scale bulk motions, which lack sufficient time to develop a cascade and can mimic a steep power spectrum (as plausibly seen in merging clusters like Coma), are thus unlikely dominant in Perseus. On the other hand, if a large fraction of the coherent bulk motion is due to sloshing circulations, the corresponding sloshing turnaround timescale in Perseus is $\sim1-5\Gyr$ at $R=100-500\kpc$, as determined by the Brunt-$\rm V\ddot{a}is\ddot{a}l\ddot{a}$ frequency \citep{Churazov2003}, which is close to the turbulence turnaround timescale in our estimation. The form of the heating rate (Eq.~\ref{eq:heating_rate}) applies to both types of motions, although the coefficient may differ. Given their comparable timescales, the presence of sloshing bulk motions would not significantly affect our estimations of the turbulent heating rate. Indeed, even in numerical simulations, it is physically non-trivial to distinguish (or even define) stratified large-scale eddies and sloshing circulations \citep[e.g.,][]{Zuhone2013}.

\section{Merger scenarios of Perseus} \label{sec:merger}

Recent multi-wavelength observations, including our XRISM measurements, suggest that the Perseus cluster is not as relaxed as previously thought. The significant velocity offset between the central ICM and the BCG \citep{XRISM2025_Perseus}, the sequence of sloshing cold fronts extending out to at least $\simeq700\kpc$ \citep{Simionescu2012,Walker2018}, and the presence of various types of extended radio emission \citep{Gendron-Marsolais2020,vanWeeren2024} all indicate a lingering memory of past (or recent) merger events. These features provide crucial insights into the merger configuration of the system.

\begin{figure*}
\centering
\includegraphics[width=0.9\linewidth]{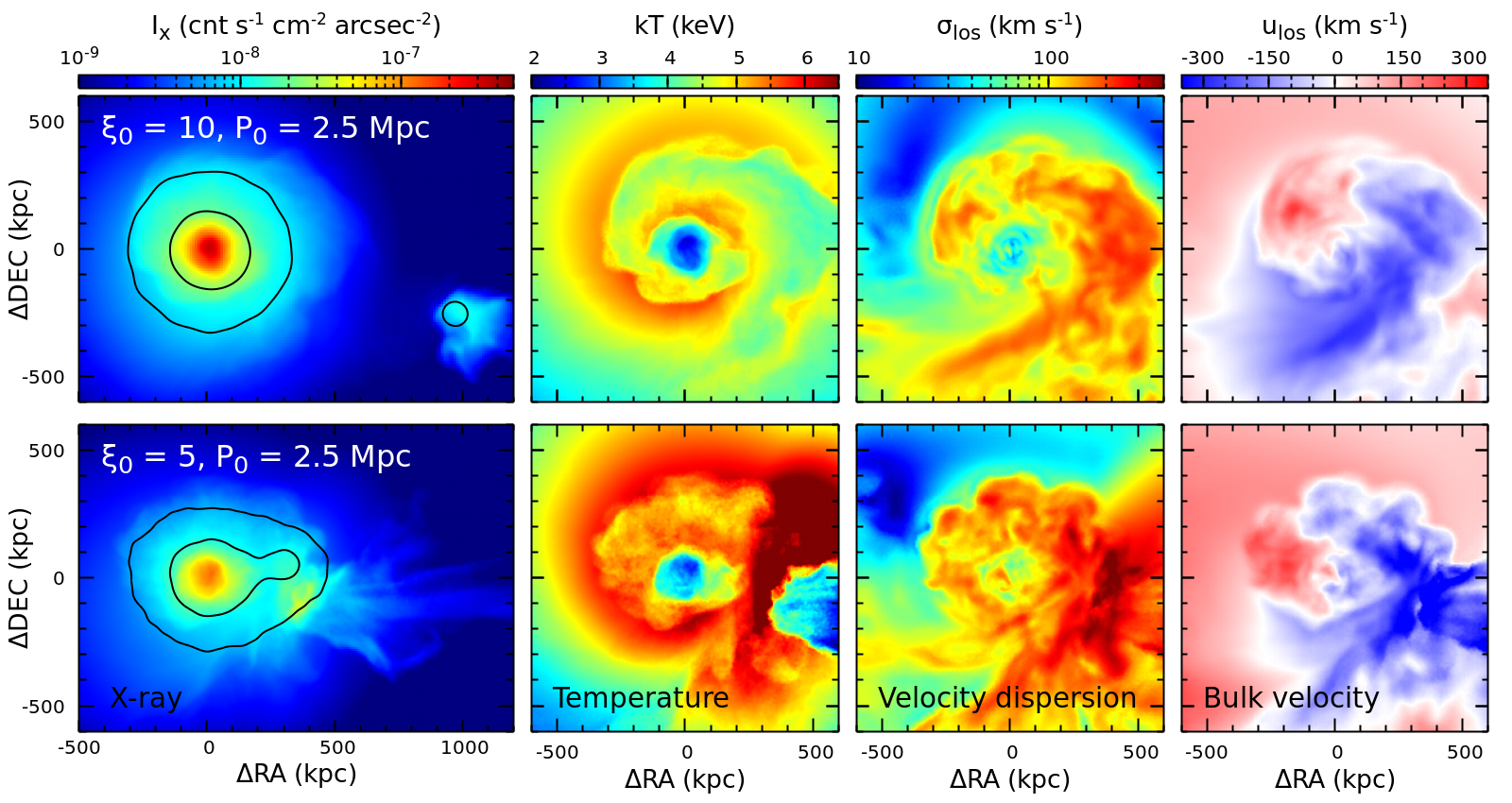}
\vspace{-5pt}
\caption{Simulated X-ray surface brightness ($0.5-8\keV$), X-ray-weighted temperature, LOS velocity dispersion, and bulk velocity for simulations with $(\xi_0,\,P_0)=(10,\,2.5\Mpc)$ and $(5,\,2.5\Mpc)$ at $t=5.0$ and $3.2\Gyr$, viewed at inclination angles $i\simeq40^\circ$, respectively. Black contours in the left panels mark the total mass distribution. The models in the top and bottom panels resemble the subcluster perturbers associated with IC310 and NGC1264, which drive the inner sloshing spiral in Perseus, respectively. Both simulations reproduce the asymmetric dipole-like structure in the bulk velocity (see Section~\ref{sec:merger:sim}). }
\label{fig:merger_mock}
\end{figure*}

\citet{Bellomi2024} investigated numerically the formation of the sloshing cold fronts and concluded that the inner ($\lesssim200\kpc$) and outer ($\simeq700\kpc$) cold fronts are unlikely to have arisen together from a single gravitational perturbation, implying either multiple pericentric passages of a single subhalo or distinct merger events. They showed in their simulations that, in the former scenario, the timescale required to develop the outer cold front is rather long ($\sim6-9\Gyr$), suggesting the subhalo may have already fully merged into the main cluster. Interestingly, two new observational studies have identified possible remnant of such merging subhalo after \citet{Bellomi2024}. \citet{HyeongHan2025} associated it with NGC1264 based on weak lensing measurements, while \citet{Churazov2025} proposed IC310 as a candidate based on its extended X-ray emission from SRG/eROSITA (see also \citealt{Schwarz1992,Furusho2001}). Both candidates are located to the west of Perseus, at projected distances of $\simeq430\kpc$ and $1000\kpc$ from the cluster center, respectively.

Our XRISM/Resolve kinematic maps offer complementary information to the dynamical structure of the Perseus Cluster. It is therefore timely to revisit the merger processes in Perseus in order to identify scenarios that can account for all these new observational features. Given the inherent complexity of cluster mergers (e.g., interaction among multiple subhalos or even filaments, non-spherical halo shapes), our goal is not to reproduce all observations in detail but to provide insights into plausible merger configurations based on hydrodynamic simulations. Our exploration is based on the following considerations, motivated by both the aforementioned observations and numerical experiments.
\begin{itemize}
  \item The merging subhalo appears to have survived west of the main cluster, particularly the candidate at $R \sim1\Mpc$, suggesting that the outer, older cold front in the east was likely generated by a separate merger a few Gyr earlier. In spite of that, our simulations focus only on the single-merger process and its resulting sloshing spiral within $\sim400\kpc$.
  \item Producing an intact, quasi-circular sloshing spiral requires an off-axis merger. A recent major merger is thus disfavored, as major mergers follow significantly more radial orbits than minor mergers based on cosmological simulations \citep[e.g.,][]{Vitvitska2002,Benson2005}; while a subhalo that is too small would fail to sufficiently disturb the atmosphere and induce sloshing \citep{Bellomi2024}. A merger mass ratio of $\sim5$–$10$ is the most plausible range.
  \item Multiple orbital passages of a subhalo tend to blur and blend the spiral structures in the cluster core. The sharp, regularly shaped cold fronts in Perseus therefore suggest that the subhalo is still on its primary orbit.
\end{itemize}

Our simulation set-ups are similar to \citet[][see also \citealt{Zhang2015} and Appendix~\ref{sec:appendix:sim} for more details]{Bellomi2024}, modeling mergers between two idealized galaxy clusters. Each cluster consists of a spherical dark matter (DM) halo and gas halo. The main cluster is tailored for Perseus to approximately capture its gas density and temperature radial profiles. The key parameters that determine the cluster merging trajectories include the merger mass ratio ($\xi_0$), initial impact parameter ($P_0$), and pairwise velocity ($V_0$). We fixed $V_0=700\kms$ in all our simulations motivated by an empirical scaling relation based on cosmological simulations \citep{Dolag2013}, and explored a broad range of $\xi_0\,(=3-10)$ and $P_0\,(=1.5-3.5\Mpc)$ for different subhalo trajectories. Each of our simulations run more than $10\Gyr$ to trace at least three pericentric passages. But none of them can produce/maintain clear sloshing spirals after the subhalo's first orbit.

\subsection{Simulation results} \label{sec:merger:sim}

Fig.~\ref{fig:merger_zplane} presents examples of our simulations: off-axis minor mergers with varying mass ratios and impact parameters, shown at snapshots after the primary apocentric passage. The turbulent wake of the infalling subhalo disrupts the development of the sloshing spiral or even rapidly destroys it by inducing instabilities and enhancing gas mixing (see the left panels). Sharp, regularly-shaped cold fronts are formed only when the wake passes by without a strong interaction with the gas core. It requires a large pericentric separation and/or a small subhalo. As shown in the middel and right panels, prominent sloshing spirals appear in the gas entropy distributions, extending to $\sim300-400\kpc$, consistent with the expansion rate of sloshing arms predicted by \citet[][$\sim80\kpc\Gyr^{-1}$]{Bellomi2024}. While magnetic fields help stabilize cold fronts, they do not substantially change the picture \citep{Bellomi2024}.

Arrows in the bottom panels of Fig.~\ref{fig:merger_zplane} show the velocity fields in the merger plane. In all cases, the gas core rotates counterclockwise, following the trajectory of the subhalo as it penetrates the main cluster. The merger with $\xi=10$ produces the most regular and least turbulent sloshing pattern, while more massive subhalos drive higher rotational velocities owing to their stronger gravitational perturbations. Despite the sparse sampling, our XRISM measurements reveal a dipole-like bulk-velocity pattern along the east–west direction with an amplitude of approximately $\pm 200-300\kms$ (see Fig.~\ref{fig:vmap}). It provides a constraint on the system's viewing direction. An inclination angle of $\simeq30^\circ-50^\circ$, defined as the angle between the LOS and $z$-axis in the simulation, is required, with a mild dependence on the merger mass ratio.

\begin{figure}
\centering
\includegraphics[width=0.95\linewidth]{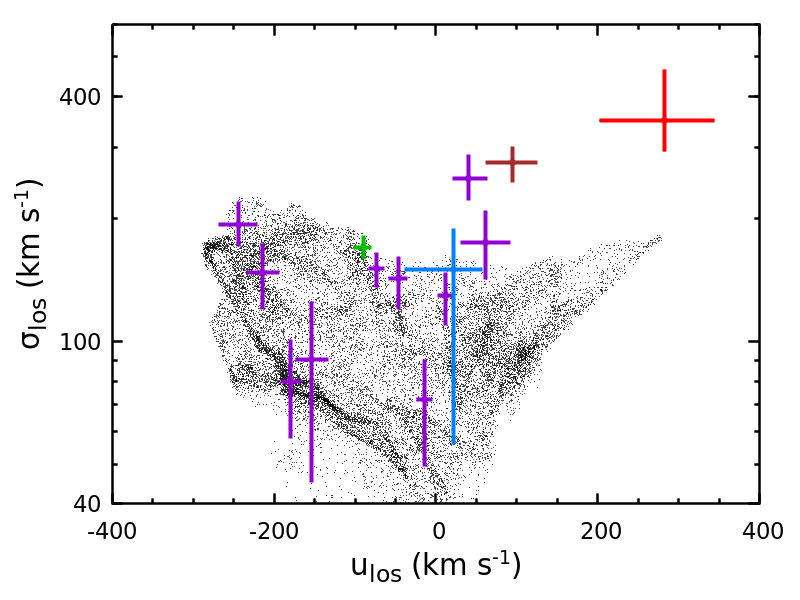}
\vspace{-5pt}
\caption{Comparison of the LOS bulk velocity -- velocity dispersion space distribution between XRISM observations and the simulation with $(\xi_0,\,P_0)=(10,\,2.5\Mpc)$ shown in the top panels of Fig.~\ref{fig:merger_mock}. Black dots mark measurements from a uniform grid within $R=400\kpc$ in the simulation. Observational data are the same as in Fig.~\ref{fig:contours}, but only $1\sigma$ errors are shown. Our single-merger simulation reproduces the overall velocity pattern but fails to capture the high velocity dispersions observed in certain outer regions of the cluster (see Section~\ref{sec:merger:sim}).}
\label{fig:vel_phase}
\end{figure}

Fig.~\ref{fig:merger_mock} shows our modeled projections with snapshots and viewing angles selected to match (1) the bulk velocity distribution and (2) the location of the subhalo as identified in \citet[][top panels]{Churazov2025} and \citet[][bottom panels]{HyeongHan2025}, respectively. Both simulations reproduce the asymmetric dipole-like velocity pattern. We did not include XRISM/Resolve's instrumental effects in the modeling, as they are not important for comparisons on $\sim100\kpc$ scales. The top panels, modeling the merger scenario suggested in \citet{Churazov2025} with an inclination angle of $i\simeq40^\circ$, resemble major observational features within $\sim400\kpc$. The subhalo's trajectory is similar to that indicated by the long radio tail of IC310 \citep{vanWeeren2024}, and its LOS velocity ($\simeq300\kms$) is consistent with the peculiar velocity of the galaxy. However, the velocity dispersion in the outer regions (around $R\simeq200-400\kpc$) is only $\sim100-200\kms$, lower than the measurements from N+NE and O3. A direct comparison of the $u_{\rm los}-\sigma_{\rm los}$ space distribution between our observation and merger simulation is shown in Fig.~\ref{fig:vel_phase}. A slightly bigger subhalo (e.g., $\xi\sim5-10$) partially alleviates this discrepancy (see the bottom panels). The cosmological environment of Perseus, together with its complex merger history, however, likely also contribute to this mismatch, which cannot be captured by our idealized single-merger simulations.

We further examine the possibility that NGC1264 and its associated subhalo are responsible for the recent sloshing in Perseus, as proposed by \citet{HyeongHan2025}. While we slightly relax the merger mass ratio inferred from weak-lensing measurements (i.e., $\xi\sim3$), a value of $\xi=5$ remains consistent within the observational uncertainties. Nevertheless, this scenario still faces two major issues that cannot be reconciled. First, a significant fraction of the subhalo’s gas atmosphere is expected to survive after the primary apocentric passage, as shown in the bottom panels of Fig.~\ref{fig:merger_mock}. Its absence in X-ray observations implies a lower gas fraction (even gasless) for the survived subhalo. Second, the peculiar velocity of NGC1264 in Perseus is $\sim-1800\kms$, comparable to the total subhalo velocity in our simulation (see Fig.~\ref{fig:merger_zplane}). Reproducing such a high LOS velocity would require a sightline nearly aligned with the subhalo’s infall direction, which contradicts the dipole-like structure (its orientation) of the ICM bulk velocity. In the projection shown in the bottom panels of Fig.~\ref{fig:merger_mock}, the subhalo’s LOS velocity is only $\sim-500\kms$. It is therefore unlikely that the subhalo identified in weak-lensing measurements is the perturber of the sloshing within $\sim400\kpc$ in Perseus. However, it can be a remnant of a previous merger that has undergone multiple orbits and lost most of its gas content, and be responsible for the eastern X-ray excess and/or the outer cold front at $\sim700\kpc$.

Taken together, we suggest that the observed features in Perseus were shaped by at least two mergers. The most recent, $\sim3-5\Gyr$ ago, produced the spiral structures within $\sim400\kpc$, with the subhalo (likely IC310) having recently passed its primary apocenter. The cold front at $R\simeq700\kpc$ was instead generated by an earlier merger, $\sim6-9\Gyr$ before, as inferred from the expansion rate of the sloshing arm \citep{Bellomi2024}. Interestingly, \citet{Zhang2020b} showed that a merger shock formed $\sim8\Gyr$ ago (around redshift $z\sim1$) is able to explain the giant contact discontinuity found near the virial radius ($\sim1.7\Mpc$) of Perseus \citep{Walker2022} through a collision between merger and accretion shocks \citep{Zhang2020a}, which may correspond to the same event that produced the $R\sim700\kpc$ cold front.

We also note that, in both models shown in Fig.~\ref{fig:merger_mock}, the LOS velocity dispersions remain consistently below $\simeq100\kms$ within the very central regions ($\lesssim60\kpc$), significantly lower than the observational measurements ($\simeq200\kms$; see Fig.~\ref{fig:vmap} and \citealt{XRISM2025_Perseus}). This suggests that a single merger alone cannot account for the central peak of the velocity dispersion in Perseus, in line with the argument in \citet{XRISM2025_Perseus} that AGN feedback plays a crucial role in boosting the observed velocity dispersion at the center.

\section{Conclusions} \label{sec:conclusion}

We present extended, up-to-date gas kinematic maps of the Perseus cluster by combining five new XRISM/Resolve pointings observed in 2025 with the PV data from 2024 \citep{XRISM2025_Perseus}. These observations cover multiple radial directions (i.e., E/NE, S, NW) and a broad range of dynamical scales ($\sim50-500\kpc$). To date, Perseus remains the only cluster that has been extensively mapped with a microcalorimeter/high spectral resolution out to $\simeq0.7r_{2500}$ (along multiple radial directions), while simultaneously offering sufficient spatial resolution to resolve gaseous substructures. This unique dataset enables an unprecedented exploration of the merger-driven gas kinematics in the ICM and provides valuable insights into the merger history of the system. Our main findings are summarized below.
\begin{itemize}
  \item We measure high velocity dispersions ($\simeq300\kms$) in the east of the cluster (see Fig.~\ref{fig:vmap}), spatially coinciding with the extended X-ray surface brightness excess seen in the high-resolution images. These velocities correspond to a nonthermal pressure fraction of $\simeq7-14\%$, in agreement with the numerical predictions of the Perseus-like clusters in the TNG-Cluster cosmological simulation. In contrast, the fractions in all other (non-AGN-dominant) regions remain consistently below $5\%$, lower than the TNG-Cluster predictions but comparable to those seen in the most relaxed clusters in TNG300 simulation suite (see Section~\ref{sec:velocity:fnth}).
  \item We characterize the velocity maps in the non-AGN-dominant region ($\gtrsim60\kpc$) using the second-order velocity structure function and the radial profile of the turbulent heating rate. The observed velocity distributions can be effectively described by a field driven by a single, large-scale kinematic source. This interpretation is supported by both the shape of the velocity structure function and the approximately uniform turbulent heating rate within $R\simeq60-400\kpc$. Our measurements favor an energy injection scale of at least a few hundred kpc (see Sections~\ref{sec:velocity:vsf} and \ref{sec:velocity:qheat}).
  \item An asymmetric dipole-like pattern along the east-west direction, with an amplitude of approximately $\pm200-300\kms$, is observed in the bulk velocity distribution (see Fig.~\ref{fig:vmap}), indicating moderate gas rotation produced by the recent merger process. The E/NE regions correspond to the receding (most positive) side, while the NW arm corresponds to the approaching (most negative) side. This feature provides a strong constraint on the LOS direction, i.e., an inclination angle of $\simeq30^\circ-50^\circ$, with a  mild dependence on the merger mass ratio.
  \item Our idealized merger simulations suggest that Perseus has undergone at least two energetic mergers since redshift $z\sim1$. The more recent event was an off-axis, minor merger with a mass ratio of $\sim5-10$, which occurred$\sim3-5\Gyr$ ago and produced the inner spiral structure ($\lesssim400\kpc$). The remnant of the infalling subcluster has recently passed its primary apocenter, likely corresponding to the radio galaxy IC310. An earlier merger, $\sim6-9\Gyr$ ago, generated the ancient cold front at $\sim700\kpc$. The remnant of that subcluster may correspond to the ``gasless'' substructure revealed by the weak-lensing measurements (see Section~\ref{sec:merger}).
\end{itemize}

This work demonstrates the power of high-resolution, non-dispersive imaging spectroscopy for probing ICM kinematics and advancing our understanding of cluster dynamics. It paves the way for future projects/missions dedicated to mapping the ICM (e.g., XRISM key projects, HUBS, and NewAthena). Nevertheless, we emphasize that the current sparse velocity coverage limits the statistical significance of our results. Additional XRISM/Resolve observations with a more uniform spatial coverage will be crucial to validate our conclusions.

\begin{acknowledgements}

CZ and NW were supported by the GACR EXPRO grant No. 21-13491X.
CZ and IZ were partially supported by NASA grant 80NSSC18K1684.
CZ, IZ, and AH acknowledge partial support from the Alfred P. Sloan Foundation through the Sloan Research Fellowship.
EM acknowledges support from NASA grants 80NSSC20K0737 and 80NSSC24K0678.
SU acknowledges support by Program for Forming Japan's Peak Research Universities (J-PEAKS) Grant Number JPJS00420230006.
This work was supported by JSPS KAKENHI grant numbers JP25K23398 (SU).
Part of this work was performed under the auspices of the U.S. Department of Energy by Lawrence Livermore National Laboratory under Contract DE-AC52-07NA27344.
The material is based upon work supported by NASA under award number 80GSFC24M0006.
The simulations presented in this paper were carried out using the Midway computing cluster provided by the University of Chicago Research Computing Center. The software used in this work was developed in part by the DOE NNSA- and DOE Office of Science supported Flash Center for Computational Science at the University of Chicago and the University of Rochester.

\end{acknowledgements}


\begin{appendix} 
\section{Supplementary Observational Details} \label{sec:appendix:obs}



\subsection{New XRISM pointings}

We include five new pointings observed after the XRISM PV phase in our analysis. Their observation IDs and pointing coordinates are listed in Table~\ref{tab:params_appendix}. The O3 pointing was observed in July 2025 with a heliocentric correction of $\simeq24\kms$ and the rest four were observed in January-February 2025 with heliocentric corrections of $\simeq-28- -25\kms$.

\begin{table}[!h]
\centering
\renewcommand{\arraystretch}{1.8}
\begin{minipage}{\linewidth}
\centering
\caption{Additional information of the calibration/GO pointings.}
\label{tab:params_appendix}
\begin{tabular}{cccc}
  \hline
  Name & ObsID & (RA,\,DEC) & Note \\
  \hline
  E  & 101017010 & (50.2814,\,41.5317) & Calibration   \\
  NE & 101018010 & (50.2638,\,41.7026) & Calibration  \\
  N  & 101019010 & (49.8244,\,41.6751) & Calibration  \\
  M3 & \makecell{ 201079010 \\ 201079020 } & (50.0000,\,41.4300) & GO (Cycle 1)  \\
  O3 & 201080010 & (50.0341,\,41.3861) & GO (Cycle 1)  \\
\hline
\hline
\vspace{-25pt}
\end{tabular}
\end{minipage}
\end{table}

\subsection{Radial profile of the Fe abundance}

\begin{figure}[!h]
\centering
\includegraphics[width=0.95\linewidth]{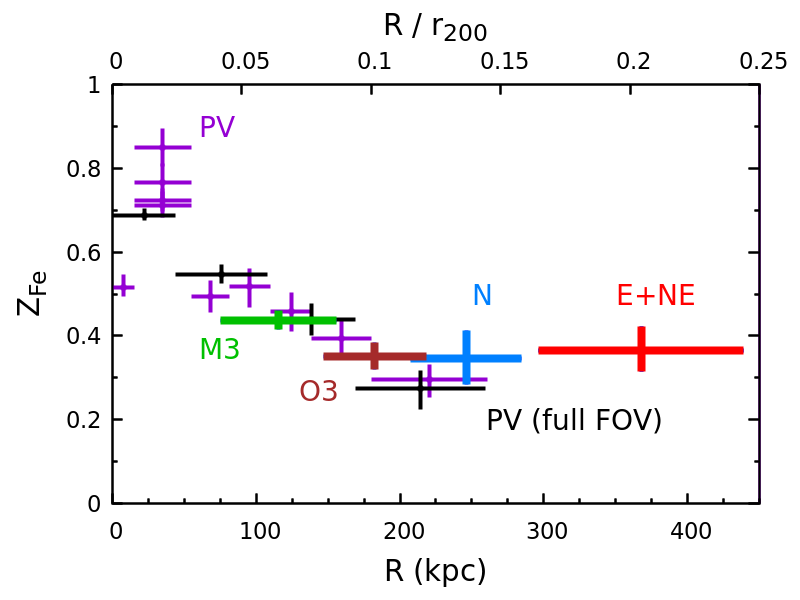}
\vspace{-5pt}
\caption{Radial profile of the Fe abundance, using the same color scheme as in Fig.~\ref{fig:contours}. Black points show the results from PV pointings, full-FOV fit \citep{XRISM2025_Perseus}. }
\label{fig:zfe}
\end{figure}

Fig.~\ref{fig:zfe} shows the radial profile of the Fe abundance -- the most extended measurements by XRISM/Resolve to date, alongside those for A2029 \citep{XRISM2025_A2029_Outer}. We emphasize that the central bright AGN and imperfect PSF calibration may introduce significant systematic uncertainties for the inner sub-regions ($\lesssim60\kpc$), preventing us from firmly concluding whether an abundance drop exists in the center. For reference, we show also the full-FOV fit of the PV data with SSM correction \citep{XRISM2025_Perseus}. The profile flattens at $\sim150\kpc$, generally consistent with previous \xmm{} and \suzaku{} measurements \citep[e.g.,][]{Churazov2003,Ueda2013,Werner2013}.

\subsection{X-ray emissivity effective length} \label{sec:appendix:obs:leff}

\begin{figure}[!h]
\centering
\includegraphics[width=0.95\linewidth]{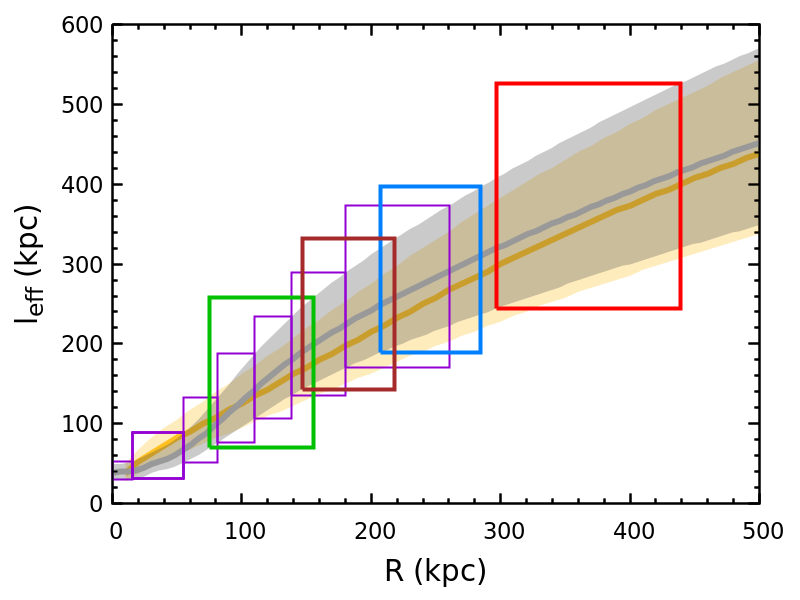}
\vspace{-5pt}
\caption{Effective length radial profiles in Perseus (grey) and A2029 (yellow). The shaded bands indicate the scales
contributing $40-60\%$ of the flux, which are adopted as the uncertainties in our $\ell_{\rm eff}$ estimates. The boxes mark the scale ranges covered by our regions in Perseus (color scheme as in Fig.~\ref{fig:contours}; see also Table~\ref{tab:params}).}
\label{fig:leff}
\end{figure}

We follow \citet{XRISM2025_Perseus} to estimate the effective length $\ell_{\rm eff}$, defined as the region size that contributes $50\%$ of the flux to the total flux at each projected distance from the cluster center. The uncertainties are characterized by varying the contributing fraction by $\pm10\%$ (shown as the shaded regions in Fig.~\ref{fig:leff}). The Perseus cluster (grey) and A2029 (yellow) exhibit comparable effective length radial profiles.

\subsection{A joint fitting between the E and NE pointings} \label{sec:appendix:obs:e+ne}

\begin{figure}[!h]
\centering
\includegraphics[width=0.95\linewidth]{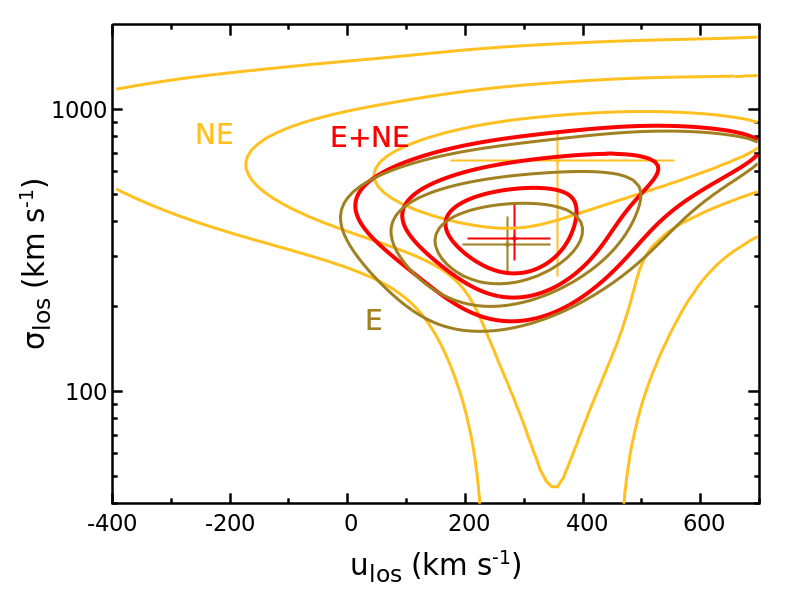}
\vspace{-5pt}
\caption{Same as Fig.~\ref{fig:contours}, but for the E and NE pointings modeled separately (orange and gray contours). The combined fit (red), included for the comparison, is the same as in Fig.~\ref{fig:contours}. }
\label{fig:contours_appendix}
\end{figure}

Fig.~\ref{fig:contours_appendix} compares the velocities measured individually from the E and NE pointings with those obtained from their joint fitting (E+NE), assuming a single-temperature model. They all exhibit consistent bulk velocities and velocity dispersions. In the joint fit, all parameters except the normalization are tied between the models for the E and NE spectra.

\subsection{Multiple components in the region E} \label{sec:appendix:obs:e2comp}

\begin{figure}[!h]
\centering
\includegraphics[width=0.95\linewidth]{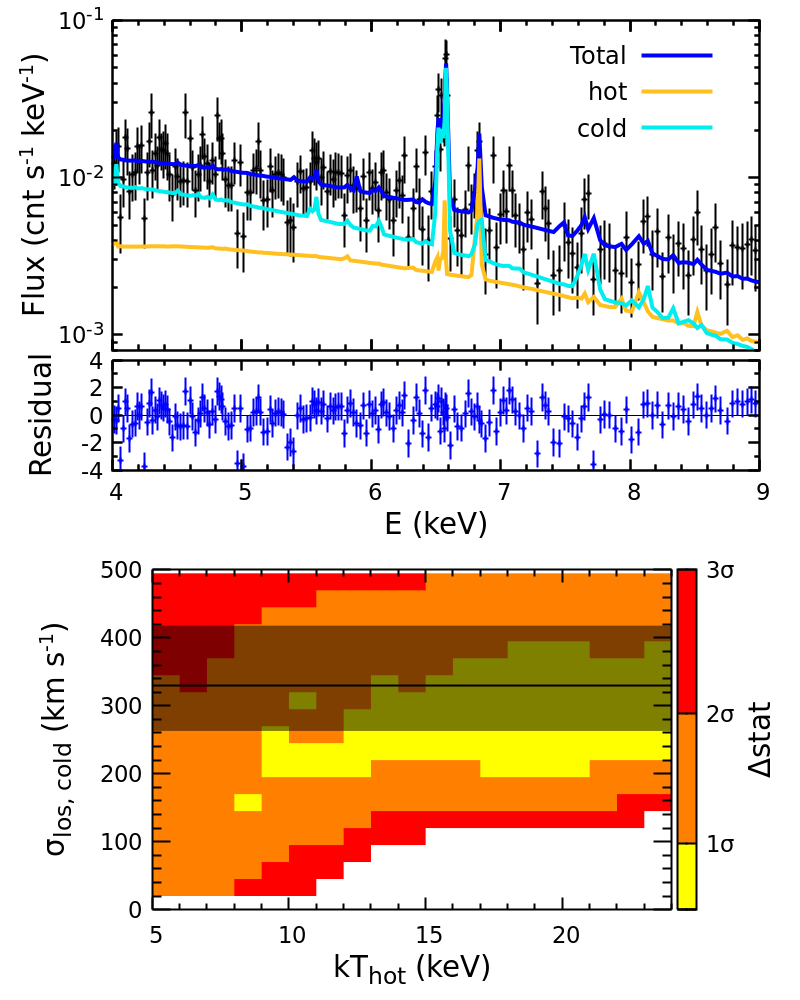}
\vspace{-5pt}
\caption{\textit{Top panels:} Spectrum in the region E with the best-fit 2T model (blue). The cyan and yellow lines show its cold and hot ICM components, respectively. \textit{Bottom panel:} $\Delta$-statistic map in the 2D parameter space: the hot component's temperature vs. the cold component's velocity dispersion. The color indicates the 1, 2, and $3\sigma$ regions. The hot component cannot be tightly constrained in our fittings.  }
\label{fig:specs_2t_appendix}
\end{figure}

A $\sim2-3\sigma$ deviation between the best-fit 1T model and the Fe~Ly$\alpha$ lines hints the presence of an additional hotter component in the region E (see Fig.~\ref{fig:e2comp}), which dominates the Fe~Ly$\alpha$ lines and is redshifted relative to the other, colder component. To explore this in more detail, we fitted the spectrum with two \texttt{bvapec} components, neglecting resonance scattering effects, as no reasonable fit could otherwise be obtained. The top panel of Fig.~\ref{fig:specs_2t_appendix} shows the best-fit 2T model along with its two individual ICM components: the hot and cold dominating the Fe~Ly$\alpha$ and He$\alpha$ line fluxes, respectively (see also Fig.~\ref{fig:e2comp} for a zoom-in near $6.7\keV$). Their relative velocity is $\simeq540\pm100\kms$. The temperature and velocity dispersion of the hot component cannot be tightly constrained. The bottom panel of Fig.~\ref{fig:specs_2t_appendix} shows the parameter space of the hot component's temperature vs. the cold component's velocity dispersion, where colors show the $\Delta$-statistic in units of $\sigma$. The velocity dispersion of the cold component (the dominant one) mildly depends on the temperature of the hot component. There are still large uncertainties in this 2T fit, limited by the quality of the current data. Deeper observations of the region are required to confirm the results.

\subsection{The \texttt{lremover} model in Xspec} \label{sec:appendix:obs:lremover}

\begin{figure}[!h]
\centering
\includegraphics[width=0.95\linewidth]{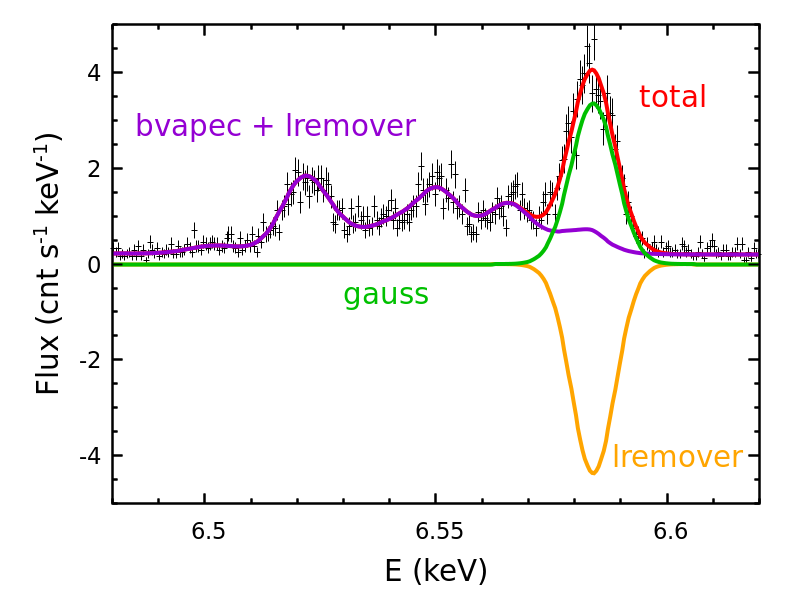}
\vspace{-5pt}
\caption{An example of using the \texttt{lremover} model to remove the Fe He$\alpha$ $w$-line from the \texttt{bvapec} model. The model generates a negative Gaussian line that exactly cancels the $w$-line in \texttt{bvapec}. }
\label{fig:lremover}
\end{figure}

Instead of manually modifying the atomic database \citep[e.g.,][]{Hitomi2018_Velocity}, we used the \texttt{lremover} model\footnote{\href{https://github.com/Congyao-Zhang/lremover}{https://github.com/Congyao-Zhang/lremover}} to remove the Fe He$\alpha$ resonance line ($w$-line) from the \texttt{bvapec} model in Xspec and replaced it with an external Gaussian component to account for the resonance scattering effect \citep[e.g.,][]{Gilfanov1987}. The main idea of the model is to introduce a negative Gaussian line that cancels the $w$-line (or any other specified line or lines) in \texttt{bvapec}. Fig.~\ref{fig:lremover} shows an example of the best-fit model, \texttt{bvapec+lremover+gauss}, for the central pointing of Perseus. The same model was employed in \citet{XRISM2025_Perseus}.

\section{Simulation methods} \label{sec:appendix:sim}

We performed two types of numerical simulations to support theoretical interpretation of the measured gas kinematic fields, including (1) turbulent simulations using the mesh-based magnetohydrodynamic code {\sc FLASH4.6} \citep{Fryxell2000} and (2) idealized cluster merger simulations using the moving-mesh code {\sc Arepo} \citep{Springel2010,Weinberger2020}.

\subsection{Turbulent simulations}

\begin{figure}[!h]
\centering
\includegraphics[width=0.95\linewidth]{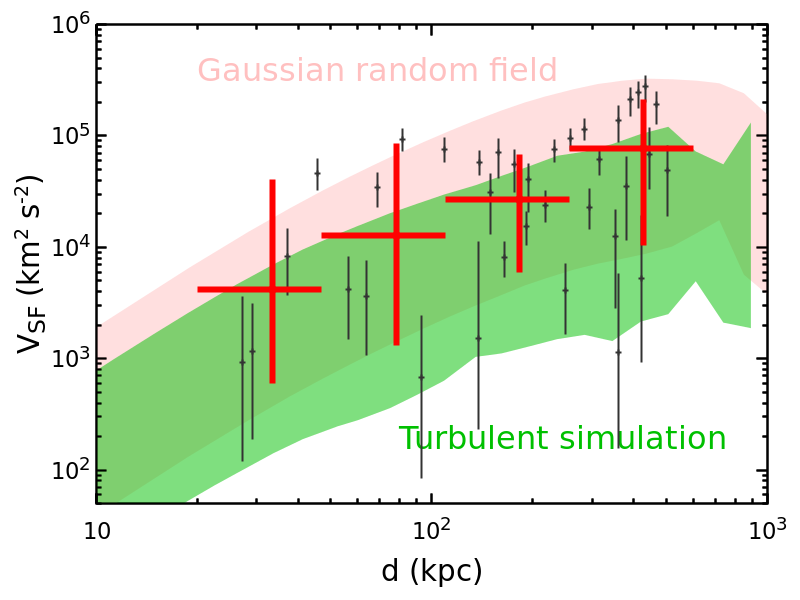}
\vspace{-5pt}
\caption{Comparison of the velocity structure function from the Gaussian random field and the turbulent simulation, both with $\ell_{\rm inj}=1\Mpc$. The observational data are identical to those in Fig.~\ref{fig:vsf}. }
\label{fig:vsf_sim}
\end{figure}

The turbulent simulations are similar to those presented in \citet{XRISM2025_Perseus}. Turbulence was driven in a Perseus-like cluster by a stochastic forcing term based on the Ornstein-Uhlenbeck process \citep[see][and references therein]{Federrath2010}, where we included only the solenoidal stirring mode, which dominates in galaxy clusters. The simulation domain is a cube of $1.2\Mpc$ with a spatial resolution of $5\kpc$, sufficient for the purpose of our experiments. We explored various injection scales $\ell_{\rm inj}=0.2$, 0.5, and $1\Mpc$. Fig.~\ref{fig:vsf_sim} compares the Gaussian random field and the turbulent simulation, both with $\ell_{\rm inj}=1\Mpc$. The shape and scatter range of their velocity structure functions are very similar, despite the presence of gravitational stratification in the simulations. We note that, although the amplitude of the velocity fields can be fine-tuned in our numerical model, it cannot be arbitrarily high because the simulation does not include radiative cooling. Excessive turbulent heating would rapidly redistribute the gas atmosphere, explaining the underestimate of the velocity structure function in our simulations. However, this is not in tension with the observations. The numerical set-ups assume a spatially and temporarily uniform heating rate, which is unlikely the case in actual clusters.

\subsection{Merger simulations}

In our merger simulations, we model mergers between two idealized galaxy clusters. Each cluster consists of both gas and DM halo components, which are spherical and in equilibrium in their initial conditions (see \citealt{Zhang2014,Zhang2015} for more details on our numerical set-ups).

The initial DM density radial profile within the virial radius follows the Navarro–Frenk–White form \citet{Navarro1997} and the concentration parameter is fixed to 4 for all halos in our simulations, motivated by recent weak-lensing measurements \citep{HyeongHan2025}. The gas density radial profile of the main cluster is tailored for Perseus, following the equation~F3 in \citet{Tang2017}. We adjust its density normalization to ensure that the averaged gas fraction within the virial radius is $f_{\rm gas}=0.12$. The radial temperature profile is determined by the hydrostatic equilibrium condition, which exhibits a $\sim100\kpc$ cool core consistent with the observations \citep{Churazov2003}. For the smaller subcluster, its density radial profile is assumed as the Burkert profile \citep{Burkert1995} with an averaged gas fraction $f_{\rm gas}=0.12$.

We fix the mass of the main cluster ($M_0 = 6\times10^{14}\,M_\odot$) in all our simulations, along with the initial pairwise velocity ($V_0 = 700{\,\rm km\,s^{-1}}$) between the two merging clusters, based on an empirical relation from cosmological simulations \citep{Dolag2013}. We explore other parameters that determine the merging halo trajectories, including the merger mass ratio ($\xi_0=5$ and $10$) and impact parameter ($P_0=1.5, 2.5, 3.5\,{\rm Mpc}$). The mergers are set to occur in the $x-y$ plane (i.e., the merger plane), with the initial cluster velocities aligned along the $y$-axis (see Fig.~\ref{fig:merger_zplane}). The mass resolutions of the DM and gas are $\simeq5\times10^8$ and $2\times10^7\msun$, respectively.

\end{appendix}

\end{document}